\documentclass{jfm}

\usepackage{psfrag}
\usepackage{graphicx}
\usepackage{natbib}
\usepackage{float}
\ifCUPmtlplainloaded \else
  \checkfont{eurm10}
  \iffontfound
    \IfFileExists{upmath.sty}
      {\typeout{^^JFound AMS Euler Roman fonts on the system,
                   using the 'upmath' package.^^J}%
       \usepackage{upmath}}
      {\typeout{^^JFound AMS Euler Roman fonts on the system, but you
                   dont seem to have the}%
       \typeout{'upmath' package installed. JFM.cls can take advantage
                 of these fonts,^^Jif you use 'upmath' package.^^J}%
      }
  \else
  \fi
\fi

\ifCUPmtlplainloaded \else
  \checkfont{msam10}
  \iffontfound
    \IfFileExists{amssymb.sty}
      {\typeout{^^JFound AMS Symbol fonts on the system, using the
                'amssymb' package.^^J}%
       \usepackage{amssymb}%
         \let\leq=\leqslant
         \let\geq=\geqslant
      }{}
  \fi
\fi

\ifCUPmtlplainloaded \else
  \IfFileExists{amsbsy.sty}
    {\typeout{^^JFound the 'amsbsy' package on the system, using it.^^J}%
     \usepackage{amsbsy}}
    {}
\fi

\usepackage{amsmath}
\usepackage{color}
\usepackage{bm}
\usepackage{epsfig}
\usepackage{subfigure}
\newsavebox{\astrutbox}
\sbox{\astrutbox}{\rule[-5pt]{0pt}{20pt}}

\newcounter{alphasect}
\def\alphainsection{0}

\let\oldsection=\section
\def\section{%
  \ifnum\alphainsection=1%
    \addtocounter{alphasect}{1}
  \fi%
\oldsection}%

\renewcommand\thesection{%
  \ifnum\alphainsection=1% 
    \Alph{alphasect}\arabic{alphasect}
  \else%
    \arabic{section}
  \fi%
}%

\newenvironment{alphasection}{%
  \ifnum\alphainsection=1%
    \errhelp={Let other blocks end at the beginning of the next block.}
    \errmessage{Nested Alpha section not allowed}
  \fi%
  \setcounter{alphasect}{0}
  \def\alphainsection{1}
}{%
  \setcounter{alphasect}{0}
  \def\alphainsection{0}
}%

\title[]{Extended Squire's transformation and its consequences on transient growth for a confined shear flow.}

\author[J. John Soundar Jerome and Jean-Marc Chomaz]%
{J.\ns J\ls o\ls h\ls n\ns S\ls o\ls u\ls n\ls d\ls a\ls r\ns J\ls e\ls r\ls o\ls m\ls e$^1$%
  \thanks{Email address for correspondence: soundar@dalembert.upmc.fr} \and \ns
J\ls e\ls a\ls n\ls -\ls M\ls a\ls r\ls c\ns C\ls h\ls o\ls m\ls a\ls z$^1$}

\affiliation{$^1$ Laboratoire d'Hydrodynamique (LadHyX), \'{E}cole Polytechnique, 91128 Palaiseau, France\\[\affilskip]}

\pubyear{2010}
\volume{650}
\pagerange{119--126}
\begin{document}

\maketitle

\begin{abstract}
The classical Squire transformation is extended to the entire eigenfunction structure of both Orr-Sommerfeld and Squire modes. For arbitrary Reynolds numbers $Re$, this transformation allows to solve the initial--value problem for an arbitrary $3D$ disturbance via a $2D$ initial--value problem at a smaller Reynolds number $Re_{2D}$. Its implications on the transient growth of arbitrary $3D$ disturbances is studied. Using the Squire transformation, the general solution of the initial--value problem is shown to predict large Reynolds number scaling for the optimal gain at all optimization time $t$ with ${t}/{Re}$ finite or large. This result  is an extension of the well-known scaling laws first obtained by \cite{Gustavsson_1991} and \cite{Reddy_n_Henningson_1993} for arbitrary $\alpha Re$, where $\alpha$ is the streamwise wavenumber. The Squire transformation is also extended to the adjoint problem and hence, the adjoint Orr-Sommerfeld and Squire modes. It is, thus, demonstrated that the long-time optimal growth of $3D$ perturbations as given by the exponential growth (or decay) of the leading eigenmode times an extra-gain representing its receptivity, may be decomposed as a product of the gains arising from purely $2D$ mechanisms and an analytical contribution representing $3D$ growth mechanisms equal to $1+ \left(\beta Re/Re_{2D}\right)^2 \mathcal{G}$, where $\beta$ is the spanwise wavenumber and $\mathcal{G}$ is a known expression. For example, when the leading eigenmode is an Orr-Sommerfeld mode, it is given by the product of respective gains from the $2D$ Orr-mechanism and an analytical expression representing the $3D$ lift-up mechanism. Whereas if the leading eigenmode is a Squire mode, the extra-gain is shown to be solely due to the $3D$ lift-up mechanism. Direct numerical solutions of the optimal gain for plane Poiseuille and plane Couette flow confirm the novel predictions of the Squire transformation extended to the initial--value problem. These results are also extended to confined shear flows in the presence of a temperature gradient.
\end{abstract}

%\begin{keywords}
%Authors should not enter keywords on the manuscript, as these must be chosen by the author during the online submission process and will then be added during the typesetting process (see http://journals.cambridge.org/data/\linebreak[3]relatedlink/jfm-\linebreak[3]keywords.pdf for the full list)
%\end{keywords}

\section{Introduction}
\label{sec:object_sq}
For over two decades now, the linear stability analysis of shear flows has followed two lines of thought, namely, modal stability analysis and non-modal stability analysis. The former considers solutions of the linearised Navier-Stokes equations ($LNS$) that grow, or decay, exponentially in time \citep{Lin_1955, Chandra_1961, Joseph_1976, Drazin_n_Reid_1981}. Whereas the later investigates the dynamics of disturbances over a finite-time horizon without assuming exponential time dependence \citep{Farrell_1988, Reddy_n_Henningson_1993, Schmid_n_Henningson_2001, Schmid_review_2007}. In the case of parallel shear flows, the celebrated Squire transformation \citep{Squire_1933} relates arbitrarily oriented three-dimensional $(3D)$ modal solution of non-zero streamwise wavenumber at a given Reynolds number to a two-dimensional $(2D)$ modal solution with the same total wavelength but zero spanwise wavenumber (hereafter referred to as spanwise disturbances) at a smaller Reynolds number. Since in the transformation the growth rate of $3D$ perturbations are smaller than that of $2D$ perturbations, it leads to the well-known Squire theorem which states that $2D$ modes are more unstable than $3D$ modes of same total wavelength, implying that the modal analysis can be restricted to only $2D$ disturbances without loss of generality.

The modal stability analysis for wall-bounded parallel shear flows predicts that $2D$ spanwise disturbances, in the form of  Tollmien-Schlichting waves, are the most unstable modes \citep{Tollmien_1928, Schlichting_1933}. Using a novel vibrating ribbon experiment, \cite{Schubauer_1947} measured and compared the growth rate of $TS$ waves with the modal stability theory. Later, \cite{Klebanoff_1962} described how the onset of $2D$ $TS$ instability waves can lead to $3D$ turbulent fluctuations. In a laminar boundary layer, they also use a vibrating ribbon to generate and follow the slow evolution of $2D$ $TS$ waves in a controlled environment. As the $TS$ wave amplitude exceeded $1\%$ of the free-stream velocity, they observed that the spanwise-uniform $TS$ waves exhibit a rapid growth of spanwise variations, thereby leading to longitudinal vortices. \cite{Herbert_1988} used the Floquet theory of secondary instability to describe the evolution of such spanwise periodic disturbances from $2D$ $TS$ waves. \cite{Bayly_Orzag_Herbet_1988, Kachanov_1994, Schlichting_2000} provide a review of the resulting transition scenario and its consequences on turbulence shear flows. On the other hand, experiments in the presence of high free-stream turbulence \citep{Morkovin_1968, Klebanoff_1971, Morkovin_1978, Morkovin_1984, Kendall_1985, Matsubara_and_Alfredsson_2001} show that transition is usually preceded by the presence of streamwise motion in the form of streaks and not via Tollmien-Schlichting ($TS$) waves as predicted by modal stability analysis. For example, \cite{Matsubara_and_Alfredsson_2001} demonstrated that a boundary layer which is subjected to free-stream turbulence levels in the range $1-6\%$ develops streamwise elongated regions of high and low streamwise velocity which lead to secondary instability and transition to turbulence. Such perturbation dynamics at the onset of transition were analysed by numerous experimental and direct numerical studies confirming this so-called \textit{bypass} transition scenario. (see the review \cite{Saric_Reed_Kerchen_2002} and references therin).
% In plane Poiseuille flow, the maximum non-dimensional growth rate based on half-width and centerline velocity is approximately $0.04$ for $2D$ spanwise disturbances and at this growth rate, however, it would take the mode about $57$ time units to grow one order of magnitude.

\cite{Ellingsen_n_Palm_1975} considered a streamwise-uniform disturbance in an inviscid shear flow to deduce that the streamwise velocity of these disturbances can grow linearly in time. They cite that it was E. H{\o}lland who originally suggested in his lecture notes that certain $3D$ disturbances can grow transiently in inviscid shear flows. \cite{Landhal_1980} generalized their result to all parallel inviscid constant density shear flows by showing that a wide range of initial infinitesimal $3D$ disturbances (in particular, those disturbances with a non-zero wall-normal velocity component) exhibit algebraic growth. \cite{Hultgren_Gustavsson_1981} were the first to consider such three-dimensional perturbations in the case of viscous parallel shear flows. They studied the temporal evolution of small $3D$ disturbances with large streamwise wavelength (\i.e. nearly streamwise-uniform) in viscous boundary layers. It was deduced that, at short-time, the streamwise perturbation velocity evolves according to inviscid initial--value problem analysed by \cite{Ellingsen_n_Palm_1975} and \cite{Landhal_1980}. Later, viscous dissipation dominates and the disturbance eventually decays. Further studies showed that such transient growth of disturbances exists in many parallel viscous shear flows. Using variational approach, \cite{Farrell_1988} computed the optimal $3D$ perturbations that give rise to the maximum possible transient growth at a given time interval. The kinetic energy of certain optimal perturbations can grow as large as $\mathcal{O}(Re^2)$ in plane Poiseuille \citep{Gustavsson_1991, Reddy_n_Henningson_1993} and plane Couette flows \citep{farrell_1993}. Depending on the initial conditions and the Reynolds number, nonlinear effects may become important during the transient growth of disturbances in these flows. \cite{Waleffe_PoF1995} proposed a self-sustaining process for turbulent shear flows consisting of finite amplitude streamwise rolls that create nonlinear streaks via transient growth and the nonlinear streaks undergo a secondary modal instability to form wall-normal vortices that, in turn, regenerate streamwise rolls via vortex tilting. It is now widely accepted that such self-sustaining processes form the basis of the so-called \textit{bypass} transition.

The process of short-time growth of disturbance kinetic energy in the absence of nonlinear effects can be associated with the non-normality of the governing linear operator \citep{Boberg_Brosa_1988, Farrell_1988, butler_1992, Reddy_n_Henningson_1993} i.e., the non-orthogonality of the associated eigenfunctions. Even though each eigenfunction may decay at its own growth rate (related to its eigenvalue), a superposition of non-orthogonal eigenfunctions may produce large transient growth before eventually decreasing at the rate of the least stable eigenfunction. Transient growth can also occur when an eigenvalue is degenerate and the operator is non-diagonal \citep{Hultgren_Gustavsson_1980, Shanthini_JFM1989}. For unbounded or semi-bounded shear flows, the continuous spectrum may also contribute to transient growth \citep{Hultgren_Gustavsson_1981}. But these cases are out of the scope of the present study, since we consider bounded shear flows wherein the spectrum is discrete; and we also assume that \textit{the spectrum to be non-degenerate since this occurs on a set of control parameters of zero measure} \citep{Schmid_n_Henningson_2001}.

The lift-up mechanism \citep{Moffatt_1967, Ellingsen_n_Palm_1975, Landhal_1980} and the Orr mechanism \citep{Orr_1907} are two such commonly identified disturbance growth phenomena in a shear flow. The lift-up mechanism is considered to be the dominant mechanism in many wall-bounded shear flows. According to the lift-up mechanism, an infinitesimal streamwise-uniform vortex superimposed on a parallel shear flow can lift-up low-speed fluid from the wall and push high-velocity fluid towards the wall until viscous dissipation becomes important at times of the order of Reynolds number $Re$. %Physically, the source of transient growth of disturbances is related to the inviscid vortex tilting process in the presence of base flow shear whereby a disturbance can feed on the base flow kinetic energy for a short time.
The Orr-mechanism is associated to the increase in disturbance kinetic energy due to an initial disturbance field that consists of spanwise-uniform vortices that are tilted against the direction of the base flow. Such a disturbance can grow by extracting the base flow kinetic energy via the Reynolds stress production term. Considering plane wave solutions for arbitrary $3D$ perturbations, \cite{farrell_1993} demonstrated that any growth in wall-normal velocity via the Orr-Mechanism can eventually lead through the lift-up mechanism to large amplification of the streamwise velocity. In a more recent study, \cite{Vitoshkin_2012} explained that $3D$ optimal growth arises when the spanwise vorticity and the $2D$ spanwise divergence field are in phase when the mean flow shear is positive and out of phase when the mean shear is negative.

In the case of confined viscous shear flows, disturbance growth over finite-time horizon (or non-modal behaviour) can be computed via an eigenfunction expansion \citep{Schmid_n_Henningson_2001, Schmid_review_2007}. In this context, the present article extends the classical Squire transformation to the wall-normal vorticity component of both the Orr-Sommerfeld and the Squire modes. The implications of this extended Squire transformation on the arbitrary initial-value problem of the $LNS$ are then explored. As a result, a large-Reynolds number transformation that relates the entire optimal gain curve of any $3D$ perturbation to a generic $2D$ problem is obtained (\S\ref{sec: consequences_InitValProb}). The extended Squire transformation and the resulting asymptotic solution to the $LNS$ at $Re \gg 1$ can be viewed as a generalization of the well-known large Reynolds number scaling laws first deduced by \cite{Gustavsson_1991} and \cite{Reddy_n_Henningson_1993} (\S\ref{subsec:DiscussionGustavssonReddyHenningson}).

\section{Governing equations}
\label{sec:govern_eqns_sq_part}
The evolution of $3D$ infinitesimal disturbances in a shear flow is governed by the $LNS$ equations with appropriate boundary conditions. For parallel shear flows that are homogeneous and infinite along streamwise ($x$-axis) and spanwise ($z$-axis) directions with base flow velocity $\vec{\textbf{\textit{U}}} = [U_0(y), 0, 0]^T$, the solution $\textbf{\textit{q}} = \left[ v\left( x,y,z,t \right), \eta\left( x,y,z,t \right) \right]^{T}$ (where $v$ and $\eta$ are wall-normal velocity and vorticity perturbation components, respectively) of the $LNS$ equations may be expanded in the so-called normal mode formulation \citep{Lin_1955, Chandra_1961, Joseph_1976, Drazin_n_Reid_1981, Schmid_n_Henningson_2001}:
\begin{equation}
	 \textbf{\textit{q}} = \int_{0}^{\infty} \int_{0}^{\infty} \tilde{\textbf{\textit{q}}} \left(y, t;  \alpha, \beta \right)  \mbox{e}^{i\alpha x + i\beta y}  d\alpha \mbox{ } d\beta,
	\label{eq:NormalMode}
\end{equation}
with the $LNS$ for each wave vector $\vec{k} = \left( \alpha, \beta \right)^{T}$ (where $\alpha$ and $\beta$ are the streamwise and spanwise wavenumbers, respectively) given by
\begin{equation}
	-\frac{\partial}{\partial t} 
		\begin{bmatrix}
			k^2-D^2	&0\\
			0	&1
		\end{bmatrix}
	\tilde{\textbf{\textit{q}}} =
 \begin{bmatrix}
				L^{O}	&0\\
				i\beta\frac{dU_0}{dy}	&L^{S}
	\end{bmatrix}
			\tilde{\textbf{\textit{q}}},
	\label{eq:LOSSQ_matrix}
\end{equation}
where $D = \frac{\partial}{\partial y}$ and $k^2 = \alpha^2 + \beta^2$. The symbols $L^{O}$ and $L^{S}$, respectively, denote the Orr-Sommerfeld and Squire operators \citep{Hultgren_Gustavsson_1980, Schmid_n_Henningson_2001}, namely,
\begin{align}
		L^{O} = i\alpha U_0 \left( k^2-D^2 \right) + i\alpha \frac{d^2U_0}{dy^2} +\frac{1}{Re} \left( k^2-D^2 \right)^2,
		\label{eq:L_direct_OS1}
\end{align}
\begin{align}
		L^{S} = i\alpha U_0 +\frac{1}{Re}\left( k^2-D^2 \right),
		\label{eq:L_direct_SQ1}
\end{align}
where $Re = {Ul}/{\nu}$ is the Reynolds number with $l$ and $U$ as the characteristic length and velocity scales for the nondimensionalization of the governing equations. For plane Poiseuille flow  and plane Couette flow, $l$ is the half-channel width ${h}/{2}$ and $U$ is the difference in velocity between the centreline and the channel wall.

When the flow is bounded in the cross-stream direction with no slip boundary conditions at the wall, the spectrum of \eqref{eq:LOSSQ_matrix} is discrete and complete \citep{Schensted_these_1961, DiPrima_Habetler_1969}. Considering the triangular form of the matrix in \eqref{eq:LOSSQ_matrix}, any solution  $\tilde{\textbf{\textit{q}}} \left( y, t; \alpha, \beta, Re \right)$ at a particular wave vector of \eqref{eq:LOSSQ_matrix} may be expressed as
\begin{align}
\tilde{\textbf{\textit{q}}} \left( y, t; \alpha, \beta, Re \right) = \sum^\infty_{j = 1} \left({A^{O}_{j} \hat{\textbf{\textit{q}}}^{O}_{j}} \mbox{e}^{-i\omega^{O}_{j}t} \right) + \sum^\infty_{j = 1} \left({A^{S}_{j} \hat{\textbf{\textit{q}}}^{S}_{j}} \mbox{e}^{-i\omega^{S}_{j}t} \right),
\label{eq:eig_expn_full}
\end{align}
where $\omega^{O}_{j}$ and $\hat{\textbf{\textit{q}}}^{O}_{j} = \left[ \hat{v}^{O}_{j}\left( y; \alpha, \beta, Re \right), \hat{\eta}^{O}_{j}\left( y; \alpha, \beta, Re \right) \right]^{T}$ are the Orr-Sommerfeld eigenvalues and eigenfunctions, respectively, with the complex frequency $\omega^{O}_{j}$ and wall-normal velocity $\hat{v}^{O}_{j}$ given by the Orr-Sommerfeld $(OS)$ equation
\begin{eqnarray}
	\left(i \omega^{O}_{j} (k^2-D^2) - L^{O}\right) \hat{v}^{O}_{j}\left( y; \alpha, \beta, Re \right)	 =	0,
\label{eq:wall_norm_velocity_eig}
\end{eqnarray}
with $\hat{v}^{O}_{j} = D\hat{v}^{O}_{j} = 0$ at the wall and the wall-normal vorticity $\hat{\eta}^{O}_{j}$ of the $OS$ eigenfunction is given by the forced Squire $(FS)$ equation:
\begin{eqnarray}
	\left(i \omega^{O}_{j} - L^{S}\right) \hat{\eta}^{O}_{j}\left( y; \alpha, \beta, Re \right)	 = i\beta\frac{dU_0}{dy}\hat{v}^{O}_{j}\left( y; \alpha, \beta, Re \right),
\label{eq:wall_norm_vorticity_eig}
\end{eqnarray}
with $\hat{\eta}^{O}_{j} = 0$ at the wall. The $FS$ equation \eqref{eq:wall_norm_vorticity_eig} has a solution only if $\omega^{O}_{j}$ is not in the spectrum of $L^{S}$. This condition is fulfilled except for a set of Reynolds number and wavenumber of zero measure \citep{Schmid_n_Henningson_2001} and these resonant cases will not be considered here. This implies, however, $\hat{\eta}^{O}_{j} = 0$ for $2D$ spanwise-uniform perturbations ($\beta = 0$). At this point, we introduce a new \textit{auxiliary} velocity variable $\hat{\hat{\eta}}^{O}_{j} = -i \hat{\eta}^{O}_{j}/\beta$ whose significance will be clear in the following sections. The corresponding forced Squire equation in terms of the $OS$ \textit{auxiliary} velocity is
\begin{eqnarray}
	\left(i \omega^{O}_{j} - L^{S}\right) \hat{\hat{\eta}}^{O}_{j}\left( y; \alpha, \beta, Re \right)	 = \frac{dU_0}{dy}\hat{v}^{O}_{j}\left( y; \alpha, \beta, Re \right),
\label{eq:auxiliary_wall_norm_vorticity_eig}
\end{eqnarray}
with $\hat{\hat{\eta}}^{O}_{j} = 0$ at the wall. This \textit{auxiliary} velocity $\hat{\hat{\eta}}^{O}_{j}$ has a non-zero solution when $\beta = 0$.

The Squire mode $\hat{\textbf{\textit{q}}}^{S}_{j} = \left[ 0, \hat{\eta}^{S}_{j}\left( y; \alpha, \beta, Re \right) \right]^{T}$ does not involve wall-normal velocity. The complex frequency $\omega^{S}_{j}$ and the wall-normal vorticity $\hat{\eta}^{S}_{j}$ are solutions of the eigenvalue problem given by the Squire $(SQ)$ equation:
\begin{eqnarray}
	\left(i \omega^{S}_{j} - L^{S}\right) \hat{\eta}^{S}_{j}\left( y; \alpha, \beta, Re \right)	 =	0,
\label{eq:wall_norm_vort_eig_sq}
\end{eqnarray}
with $\hat{\eta}^{S}_{j} = 0$ at the wall. The coefficients $\mbox{\{}A^{O}_{j}\mbox{\}}$ and $\mbox{\{}A^{S}_{j}\mbox{\}}$ in \eqref{eq:eig_expn_full} are determined from the initial condition.

\section{The extended Squire transformation on the eigenfunctions}
\label{sec: consequences_eigen_fn_sqtrans}
For the perturbations with non-zero streamwise wavenumber $\alpha \neq 0$, the $OS$ and $SQ$ eigenvalue problem  \eqref{eq:wall_norm_velocity_eig} and \eqref{eq:wall_norm_vort_eig_sq} are invariant under the Squire transformation which keeps the wave-vector modulus $k$ constant: $\alpha \rightarrow \alpha'$, $\beta \rightarrow \beta' = \sqrt{k^{2} - \alpha'^{2}}$,  $Re \rightarrow Re' = \left( {\alpha}/{\alpha'} \right) Re$ and $\omega \rightarrow \omega' = \left( {\alpha'}/{\alpha} \right) \omega$. Thus, for the $OS$-modes, $\hat{v}^{O}_{j} \rightarrow \hat{v}^{O'}_{j} = \hat{v}^{O}_{j}$ and $\hat{\hat{\eta}}^{O}_{j} \rightarrow \hat{\hat{\eta}}^{O'}_{j} = \left( \alpha / \alpha'\right) \hat{\hat{\eta}}^{O}_{j}$ and for the $SQ$-modes, $\hat{\eta}^{S}_{j} \rightarrow \hat{\eta}^{S'}_{j} = \hat{\eta}^{S}_{j}$. By setting $\alpha' = k$, $\beta$ vanishes and any $3D$ eigenmode is related to a $2D$ spanwise eigenmode at a smaller Reynolds number $Re_{2D} = \left( \alpha/k \right) Re$ with a larger frequency and growth rate given by $\omega_{2D} = \left( k/\alpha \right) \omega$. Implications of the classical Squire transformation are well-known for the wall-normal velocity component $\hat{v}^{O}_{j}$ of the $OS$-mode:
\begin{eqnarray}
\hat{v}^{O}_{j}(y; \alpha, \beta, Re) = \hat{v}^{O 2D}_{j}(y; k, Re_{2D}),
\label{eq:sq_transform_v}
\end{eqnarray}
where $\hat{v}^{O 2D}_{j}$ is the solution of the $2D$ Orr-Sommerfeld equation \citep{Lin_1955, Chandra_1961, Joseph_1976, Drazin_n_Reid_1981, Schmid_n_Henningson_2001}
\begin{eqnarray}
\left[i \left(\omega^{O2D}_{j} - k U_0 \right) \left( k^2 - D^2 \right) -i k \frac{d^{2}U_{0}}{dy^{2}} -\frac{1}{Re_{2D}}\left( k^2-D^2 \right)^{2} \right]\hat{v}^{O 2D}_{j}(y; k, Re_{2D})\\ \notag	 = 0,
\label{eq:2D_direct_OS}
\end{eqnarray}
with $\hat{v}^{O 2D}_{j} = D\hat{v}^{O 2D}_{j} = 0$ at the wall. However, to the authors' best knowledge, the transformation of the wall-normal vorticity component of the $OS$ and $SQ$ eigenmodes have never been considered before. Most of the results presented here are precisely due to this extension of the classical Squire transformation.

For the $OS$-mode the wall-normal vorticity $\hat{\eta}^{O}_{j}$ vanishes for the $2D$ case but the Squire transformation suggests to rewrite it in terms of the \textit{auxiliary} velocity variable as
\begin{eqnarray}
\hat{\eta}^{O}_{j}(y; \alpha, \beta, Re) = i \beta \frac{k}{\alpha} \hat{\hat{\eta}}^{O 2D}_{j}(y; k, Re_{2D}),
\label{eq:sq_transform_eta}
\end{eqnarray}
where $\hat{\eta}^{O 2D}_{j}$ is the solution of the $2D$ Squire equation forced at $\omega^{O2D}_{j}$:
\begin{eqnarray}
\left[i \left(\omega^{O2D}_{j} - k U_0 \right) - \frac{1}{Re_{2D}}\left( k^2-D^2 \right) \right]\hat{\hat{\eta}}^{O 2D}_{j}(y; k, Re_{2D})	 \\ \notag = \frac{dU_0}{dy}\hat{v}^{O2D}_{j}(y; k, Re_{2D}),
\label{eq:2D_direct_SQ}
\end{eqnarray}
with $\hat{\hat{\eta}}^{O 2D}_{j} = 0$ at the wall. Applying the Squire transformation also to the wall-normal vorticity $\hat{\eta}^{O}_{j}$ is somehow unusual, since $\hat{\eta}^{O}_{j}$ is zero in the strictly $2D$ case. However, the \textit{auxiliary} velocity variable is \textit{non-zero} when $\beta = 0$. As a result, it can be shown (see section \S\ref{sec: TransformationInPrimitiveVariables}) that the corresponding $2D$ velocity field is equivalent to a three--component $2D$ flow: three non-zero velocity components which are uniform in the spanwise direction. On the other hand, if $3D$ perturbations that are asymptotic to the longitudinal case are considered by taking $\alpha \rightarrow 0$ at constant $k$ and $Re_{2D}$ (i.e. assuming that the flow Reynolds number $Re = k \mbox{ } Re_{2D}/\alpha$ goes to infinity), equation \eqref{eq:sq_transform_eta} then implies that, the wall-normal vorticity $\hat{\eta}^{O}_{j}$ of the $OS$-mode diverges as $\alpha^{-1}$ while the wall-normal velocity $\hat{v}^{O}_{j}$ remains constant. This is another manifestation of the lift-up mechanism \citep{Moffatt_1967, Ellingsen_n_Palm_1975, Landhal_1980, Boberg_Brosa_1988, Gustavsson_1991, butler_1992, farrell_1993} in $3D$ $OS$-modes whereby the wall-normal vorticity $\hat{\eta}^{O}_{j}$ is a forced response due to the tilting of the base flow shear ${dU_{0}}/{dy}$ by the wall-normal velocity $\hat{v}^{O}_{j}$ solution of the $OS$ equation.

Similarly, for the $SQ$-mode the wall-normal vorticity $\hat{\eta}^{S}_{j}$ should vanish for the strictly $2D$ case. But if, instead, one considers the so-called three-component $2D$ flows wherein the spanwise velocity $\hat{\mbox{w}}$ is non-zero but uniform in the spanwise direction, the wall-normal vorticity is then non-zero in the $2D$-case and it corresponds to the variation of the spanwise velocity $\hat{\mbox{w}}^{S 2D}$ in the streamwise direction given by $\hat{\eta}^{S 2D}_{j} = -i k\hat{\mbox{w}}^{S 2D}$. Then, the extended Squire transformation also applies to the Squire mode with
\begin{eqnarray}
\hat{\eta}^{S}_{j}(y; \alpha, \beta, Re) = \hat{\eta}^{S 2D}_{j}(y; k, Re_{2D}),
\label{eq:sq_transform_etaSquire}
\end{eqnarray}
where $\hat{\eta}^{S 2D}_{j}$ is the $2D$ Squire eigenfunction solution of the $2D$ Squire equation valid for the three-component $2D$ flow:
\begin{eqnarray}
\left[i \left(\omega^{S2D}_{j} - k U_0 \right) - \frac{1}{Re_{2D}}\left( k^2-D^2 \right) \right]\hat{\eta}^{S 2D}_{j}(y; k, Re_{2D}) = 	0,
\label{eq:2D_direct_etaSquire}
\end{eqnarray}
with $\hat{\eta}^{S 2D}_{j} = 0$ at the wall. Equations \eqref{eq:sq_transform_eta} and \eqref{eq:sq_transform_etaSquire} relating $\hat{\eta}^{O}_{j}$ and $\hat{\eta}^{S}_{j}$, respectively, to the presently introduced $\hat{\hat{\eta}}^{O 2D}_{j}$ and $\hat{\eta}^{S 2D}_{j}$ define the extended Squire transformation.

\section{The extended Squire transformation in primitive variables}
\label{sec: TransformationInPrimitiveVariables}

It is interesting to rewrite the extended Squire transformation in terms of the normal modes of the Fourier--transformed primitive variables, namely, the streamwise velocity $\hat{u}(y; \alpha, \beta, Re)$, the wall-normal velocity $\hat{v}(y; \alpha, \beta, Re)$, the spanwise velocity $\hat{\mbox{w}}(y; \alpha, \beta, Re)$ and the pressure field $\hat{p}(y; \alpha, \beta, Re)$. In this case, the non-dimensional governing equations of the perturbation velocity and pressure field are
\begin{align}
i \alpha \hat{u} + D\hat{v} + i \beta\hat{\mbox{w}} = 0,
\label{eq:continuity_FourTransf}
\end{align}
\begin{align}
\left[ i \left( \omega - \alpha U_{0}\right) + \frac{1}{Re} \left(D^2 -k^2\right) \right] \hat{u} = i\alpha \hat{p} + \hat{v}\frac{dU_{0}}{dy},
\label{eq:xmomentum_FourTransf}
\end{align}
\begin{align}
\left[ i \left( \omega - \alpha U_{0}\right) + \frac{1}{Re} \left(D^2 -k^2\right) \right] \hat{v} = D\hat{p},
\label{eq:ymomentum_FourTransf}
\end{align}
and
\begin{align}
\left[ i \left( \omega - \alpha U_{0}\right) + \frac{1}{Re} \left(D^2 -k^2\right) \right] \hat{\mbox{w}} = i\beta \hat{p},
\label{eq:zmomentum_FourTransf}
\end{align}
with $\hat{u} = \hat{v} = \hat{\mbox{w}} = \hat{p} = 0$ at the wall. The classical Squire transformation should be valid for the primitive variables as well. Thus, for each $3D$ normal mode ($\hat{u}$, $\hat{v}$, $\hat{\mbox{w}}$, $\hat{p}$), there exists a $2D$ spanwise-uniform normal mode at a smaller Reynolds number $Re_{2D} = \left( \alpha/k \right) Re$ with a larger frequency and growth rate given by $\omega_{2D} = \left( k/\alpha \right) \omega$. It can be verified that the following extended Squire's transformation for the primitive variables exists, for all $\alpha$, $\beta$ and $Re$:

\begin{align}
\hat{u}(y; \alpha, \beta, Re) = \frac{k}{\alpha} \left[ \hat{u}^{2D}(y; k, Re_{2D}) - \frac{\beta^2}{k^2} \hat{\mbox{w}}^{2D}(y; k, Re_{2D}) \right],
\label{eq:u_sqtrans}
\end{align}
\begin{align}
\hat{v}(y; \alpha, \beta, Re) = \hat{v}^{2D}(y; k, Re_{2D}),
\label{eq:v_sqtrans}
\end{align}
\begin{align}
\hat{\mbox{w}}(y; \alpha, \beta, Re) = \frac{\beta}{k}\hat{\mbox{w}}^{2D}(y; k, Re_{2D}),
\label{eq:w_sqtrans}
\end{align}
and
\begin{align}
\hat{p}(y; \alpha, \beta, Re) = \frac{\alpha}{k}\hat{p}^{2D}(y; k, Re_{2D}),
\label{eq:p_sqtrans}
\end{align}
where the equations corresponding to the $2D$ spanwise-uniform fields are
\begin{align}
i k \hat{u}^{2D} + D\hat{v}^{2D} = 0,
\label{eq:2Dcontinuity_FourTransf}
\end{align}
\begin{align}
\left[ i \left( \omega^{2D} - k U_{0}\right) + \frac{1}{Re_{2D}} \left(D^2 -k^2\right) \right] \hat{u}^{2D} = i k \hat{p}^{2D} + \hat{v}^{2D}\frac{dU_{0}}{dy},
\label{eq:x2Dmomentum_FourTransf}
\end{align}
\begin{align}
\left[ i \left( \omega^{2D} - k U_{0}\right) + \frac{1}{Re_{2D}} \left(D^2 -k^2\right) \right] \hat{v}^{2D} = D\hat{p}^{2D},
\label{eq:y2Dmomentum_FourTransf}
\end{align}
and
\begin{align}
\left[ i \left( \omega^{2D} - k U_{0}\right) + \frac{1}{Re_{2D}} \left(D^2 -k^2\right) \right] \hat{\mbox{w}}^{2D} = i k \hat{p}^{2D},
\label{eq:z2Dmomentum_FourTransf}
\end{align}
with $\hat{u}^{2D} = \hat{v}^{2D} = \hat{\mbox{w}}^{2D} = \hat{p}^{2D} = 0$ at the wall. Equations \eqref{eq:2Dcontinuity_FourTransf}--\eqref{eq:y2Dmomentum_FourTransf} are the commonly known Squire--transformed $2D$--equivalent of equations \eqref{eq:continuity_FourTransf}--\eqref{eq:ymomentum_FourTransf} for the streamwise and wall-normal velocity components. The Squire transformation for the $\hat{u}$-component (\eqref{eq:u_sqtrans}) shows a complex behaviour related to the contributions from the $2D$ streamwise and spanwise velocity components with different scalings. Together with  the transformation for $\hat{\mbox{w}}$ (\eqref{eq:w_sqtrans}) and  the evolution equation of $\hat{\mbox{w}}^{2D}$, they can be considered as an extension to the classical Squire transformation equations. In this way, every $3D$ perturbation field can be related to a three--component $2D$ perturbation field.

The $2D$ spanwise velocity $\hat{\mbox{w}}^{2D}$ is, by definition, independent of $\beta$. As $\beta \rightarrow 0$ ($\alpha \rightarrow k$),  from \eqref{eq:u_sqtrans} we obtain that $\hat{u} \rightarrow \hat{u}^{2D}$ and from \eqref{eq:w_sqtrans}, we get,
\begin{align}
\lim_{\beta \rightarrow 0} \frac{\hat{\mbox{w}}}{\beta} = \dfrac{\hat{\mbox{w}}^{2D}}{k}.
\label{eq:w2Dlimit}
\end{align}

If $[ \hat{u}^{S2D}_{j}, \hat{v}^{S2D}_{j},$ $\hat{\mbox{w}}^{S2D}_{j}, \hat{p}^{S2D}_{j} ]^{T}$ denotes the $2D$ $SQ$--mode in primitive variables, the wall-normal velocity $\hat{v}^{S2D}_{j}$ is zero for the $2D$ $SQ$--mode; its streamwise velocity $\hat{u}^{S2D}_{j}$ and pressure field $\hat{p}^{S2D}_{j}$ should also be zero, according to equations \eqref{eq:2Dcontinuity_FourTransf}--\eqref{eq:y2Dmomentum_FourTransf}. Therefore, $2D$ $SQ$--mode in terms of the primitive variables is $[0, 0, \hat{\mbox{w}}^{S2D}_{j}, 0 ]^{T}$ which corresponds simply to a pressure--less $2D$ perturbation field with \textit{only} a spanwise velocity. This \textit{non-zero} spanwise velocity component is uniform in the spanwise direction but varies along the streamwise and wall-normal directions.

If $[ \hat{u}^{O2D}_{j}, \hat{v}^{O2D}_{j}, \hat{\mbox{w}}^{O2D}_{j}, \hat{p}^{O2D}_{j} ]^{T}$ denotes the $2D$ $OS$--mode in primitive variables, the wall-normal vorticity of any $OS$--mode is then
\begin{align}
\hat{\eta}^{O}_{j} = i \beta \frac{k}{\alpha} \left(\hat{u}^{O2D}_{j} - \hat{\mbox{w}}^{O2D}_{j}  \right) = i \beta \frac{k}{\alpha} \hat{\hat{\eta}}^{O2D}_{j}
\label{eq:etaOS_primitive}
\end{align}
in accordance with \eqref{eq:sq_transform_eta}.

Indeed, by definition, the $OS$ wall-normal vorticity $\hat{\eta}^{O}_{j}$ is given by $\hat{\eta}^{O}_{j} = i \beta \hat{u}^{O}_{j} - i \alpha \hat{\mbox{w}}^{O}_{j}$. The \textit{auxiliary} velocity variable $\hat{\hat{\eta}}^{O}_{j}$ introduced in the previous section is then
\begin{align}
\hat{\hat{\eta}}^{O}_{j}(y; \alpha, \beta, Re) = \hat{u}^{O}_{j}(y; \alpha, \beta, Re)  - \frac{\alpha}{\beta} \hat{\mbox{w}}^{O}_{j}(y; \alpha, \beta, Re),
\label{eq:etaauxi_primitive}
\end{align}
which can be rewritten using the extended Squire transformation \eqref{eq:u_sqtrans} and \eqref{eq:w_sqtrans} as
\begin{align}
\hat{\hat{\eta}}^{O}_{j}(y; \alpha, \beta, Re) = \frac{k}{\alpha} \left[\hat{u}^{O2D}_{j}(y; k, Re_{2D}) - \hat{\mbox{w}}^{O2D}_{j}(y; k, Re_{2D})\right],
\label{eq:eta2Dauxi_primitive}
\end{align}
showing that
\begin{align}
\hat{\hat{\eta}}^{O2D}_{j}(y; \alpha, \beta, Re) = \hat{u}^{O2D}_{j}(y; k, Re_{2D}) - \hat{\mbox{w}}^{O2D}_{j}(y; k, Re_{2D}).
\label{eq:eta2Dauxi_primitive2}
\end{align}
This implies that the $2D$ \textit{auxiliary} velocity variable represents the difference between the $2D$ streamwise and spanwise velocity components.

\section{The extended Squire transformation on the initial--value problem}
\label{sec: consequences_InitValProb}
The difference in the scaling of $\hat{v}^{O}_{j}$, $\hat{\eta}^{O}_{j}$ and $\hat{\eta}^{S}_{j}$ when applying the extended Squire transformation implies that the general solution \eqref{eq:eig_expn_full} to the initial--value problem \eqref{eq:LOSSQ_matrix} with the same initial condition $\tilde{\textbf{\textit{q}}}_{0}$ for various $\alpha$, $\beta$ and $Re$ corresponding to the same $Re_{2D}$ and $k$, can be rewritten as
\begin{align}
\tilde{\textbf{\textit{q}}}\left(y, t; \alpha, \beta, Re\right) =
\sum^\infty_{j = 1}{
	A^{O}_{j}
	\begin{bmatrix}
		\hat{v}^{O2D}_{j}\left(y; k, Re_{2D}\right)\\
		\left( \frac{i \beta Re}{Re_{2D}} \right)\hat{\hat{\eta}}^{O 2D}_{j}\left(y; k, Re_{2D}\right)\\
	\end{bmatrix}
	\exp{\left( -i Re_{2D}\omega^{O2D}_{j} \frac{t}{Re} \right)}
}
\notag \\+
\sum^\infty_{j = 1}{
	 \left(\frac{Re}{Re_{2D}}B^{O}_{j} + B^{S}_{j}\right)
	\begin{bmatrix}
	0\\
	\hat{\eta}^{S2D}_{j}\left(y; k, Re_{2D}\right)\\
	\end{bmatrix}}
	\exp{\left( -i  Re_{2D} \omega^{S2D}_{j} \frac{t}{Re}  \right)}.
\label{eq:eig_expn_full_with_SQ_transform}
\end{align}
Here, $A^{O}_{j}$, $B^{O}_{j}$ and $B^{S}_{j}$ are constants and depend only on the initial condition $\tilde{\textbf{\textit{q}}}_{0}$ for a given $k$ and $Re_{2D}$. Since the Squire modes do not contribute to the disturbance wall-normal velocity, the $v$-component of the initial--value $\tilde{\textbf{\textit{q}}}_{0}$, namely, $\tilde{v}_{0}$  determines the coefficients $A^{O}_{j}$ of the $OS$-modes:
\begin{align}
\sum^\infty_{j = 1}
{
	A^{O}_{j}\hat{v}^{O2D}_{j}
} = \tilde{v}_{0}.
\label{eq:coeff_vel}
\end{align}
Consequently, the coefficients $A^{S}_{j}$ of the Squire modes play a two-fold role:
\begin{enumerate}
\item a part of $A^{S}_{j}$ should cancel the wall-normal vorticity contribution from the $OS$-mode and scale as ${Re} / {Re_{2D}}$, i.e.
\begin{align}
\sum^\infty_{j = 1}{B^{O}_{j} \hat{\eta}^{S2D}_{j}} = -i \beta\sum^\infty_{j = 1}{A^{O}_{j} \hat{\hat{\eta}}^{O2D}_{j}},
\label{eq:blaahhha}
\end{align}
which is non-zero as long as $\beta \ne 0$.
\item the other part of $A^{S}_{j}$ should contribute to the initial wall-normal vorticity field $\tilde{\eta}_{0}$ of $\tilde{\textbf{\textit{q}}}_{0}$
\begin{align}
\sum^\infty_{j = 1}
{
	 B^{S}_{j}\hat{\eta}^{S2D}_{j}
} = \tilde{\eta}_{0}.
\label{eq:coeff_eta}
\end{align}
\end{enumerate}

This may be proved by considering a given $Re_{2D}$ and $k$, as $Re$ changes. For $t \ll Re/Re_{2D}$, the short-time expansion of wall-normal vorticity in the solution \eqref{eq:eig_expn_full_with_SQ_transform} gives
\begin{align}
\tilde{\eta}\left(y, t; k, Re_{2D}\right)
= \Pi_0 + k \Pi_1 t -i Re_{2D} \Pi_2 \frac{t}{Re} + \mathcal{O}\left(t^{2}\right),
\label{eq:eta_eig_expn_full_with_SQ_transform}
\end{align}
where,
\begin{align}
\Pi_0 = \frac{Re}{Re_{2D}}\sum^\infty_{j = 1}{ \left( i \beta A^{O}_{j} \hat{\hat{\eta}}^{O2D}_{j} + B^{O}_{j} \hat{\eta}^{S2D}_{j} \right)} + \sum^\infty_{j = 1}{ B^{S}_{j} \hat{\eta}^{S2D}_{j} }, 
\label{eq:eta_eig_expn_full_with_SQ_transform_Pi0}
\end{align}
\begin{align}
\Pi_1 = \sum^\infty_{j = 1}{ \left( \frac{\beta}{k} A^{O}_{j} \hat{\hat{\eta}}^{O2D}_{j} \omega^{O2D}_{j} - \frac{i}{k} B^{O}_{j} \hat{\eta}^{S2D}_{j} \omega^{S2D}_{j} \right)}, 
\label{eq:eta_eig_expn_full_with_SQ_transform_Pi1}
\end{align}
and
\begin{align}
\Pi_2 = \sum^\infty_{j = 1}{ B^{S}_{j} \hat{\eta}^{S2D}_{j} \omega^{S2D}_{j} }.
\label{eq:eta_eig_expn_full_with_SQ_transform_Pi2}
\end{align}
Since $\tilde{\eta}_{0}(y)$ is assumed to be the same for all $Re$,
\begin{align}
\sum^\infty_{j = 1}{ \left( i \beta A^{O}_{j} \hat{\hat{\eta}}^{O2D}_{j} + B^{O}_{j} \hat{\eta}^{S2D}_{j} \right)} = 0, 
\label{eq:InitEta1}
\end{align}
and hence,
\begin{align}
\sum^\infty_{j = 1}{ B^{S}_{j} \hat{\eta}^{S2D}_{j} } = \tilde{\eta}_{0}, 
\label{eq:InitEta2}
\end{align}
showing that the initial vorticity $\tilde{\eta}_{0}$ is only spanned by the Squire modes $\hat{\eta}^{S2D}_{j}$.

As $Re$ becomes very large, the leading term for $1 \ll t \ll {Re}/{Re_{2D}}$ is
\begin{align}
\tilde{\eta}(y, t; k, Re_{2D}) \sim k \Pi_1 t,
\label{eq:EtaLeadingTerms}
\end{align}
which offers the possibility for short-time growth even if $\omega^{O}_{j}$ and $\omega^{S}_{j}$ are all stable with negative imaginary parts. Since the kinetic energy of the disturbance $E_{k}$, in terms of wall-normal velocity and vorticity, reads
\begin{align}
	E_{k}(t) = \frac{1}{2} \int_{-1}^{1} \left[ \left| \tilde{v} \right|^2 + k^{-2} \left( \left| D\tilde{v} \right|^2 + \left| \tilde{\eta} \right|^2\right) \right] dy,
	\label{eq: norm_KE}
\end{align}
where, with no loss of generality, the $y$-domain is assumed to be bounded by $y = \pm 1$ for convenience. For $1 \ll t \ll {Re}/{Re_{2D}}$, the energy is led by the $\eta$-term in \eqref{eq: norm_KE}, giving:
\begin{align}
	E_{k}(t) \sim \frac{ t^{2}}{2} \int_{-1}^{1} \left| \Pi_1 \right|^{2} dy,
	\label{eq: norm_KELeadByEta}
\end{align}
The optimal growth is obtained by solving for an initial disturbance that would give rise to the maximum possible growth at a particular time horizon $t$ and it is defined by the gain function
\begin{align}
G\left(t; \alpha, \beta, Re\right) = \operatorname*{sup}_{\forall E_{k}(0) \neq  0} \frac{E_{k}(t)}{E_{k}(0)},
\label{eq:trans_gwth_defn}
\end{align}
where $E_{k}(0)$ is the initial perturbation kinetic energy. For fixed $Re_{2D}$ and $k$, the intermediate time asymptotics at $1 \ll t \ll {Re}/{Re_{2D}}$, for $Re$ going to infinity gives
\begin{align}
G\left(t; \alpha, \beta, Re\right) \sim \left( \frac{Re}{Re_{2D}} \right)^2 t_{2D}^{2} \mathcal{G}_{2D}\left(k, Re_{2D}\right),
\label{eq:G_asymp_largeRe}
\end{align}
with $t_{2D} = t {Re_{2D}}/{Re}$ and
\begin{align}
\mathcal{G}_{2D} \left(k, Re_{2D}\right) = \operatorname*{sup}_{\forall E_{k}(0) \neq 0} \left[ \frac{\frac{1}{2} \int_{-1}^{1} \left| \Pi_1 \right|^{2} dy}{E_{k}(0)} \right],
\label{eq:G2D}
\end{align}
which is a function of $k$ and $Re_{2D}$, independent of time and Reynolds number $Re$, since $\Pi_1$ depends only on $A^{O}_{j}$, $\hat{\hat{\eta}}^{O2D}_{j}$, $\omega^{O2D}_{j}$, $B^{O}_{j}$, $\hat{\eta}^{S2D}_{j}$ and $\omega^{S2D}_{j}$. Furthermore, $\Pi_1$ is independent of $B^{S}_{j}$ and maximizing $\mathcal{G}_{2D}$ then imposes $B^{S}_{j} = 0$ which gives $\tilde{\eta}_{0}(y) = 0$. Thus, the optimal in \eqref{eq:G2D} should be looked for within initial conditions on $\tilde{v}_{0}(y)$ only.

For time $t \gtrsim Re/Re_{2D}$, the large Reynolds number $Re$ asymptotics for the energy is given by
\begin{align}
	E_{k}(t) \sim  \left( \frac{Re}{Re_{2D}} \right)^2 \mathcal{I}_{2D} \left( t_{2D}; k, Re_{2D} \right),
	\label{eq: norm_KELeadByEta_largeRe}
\end{align}
where
\begin{align}
\mathcal{I}_{2D} \left( t_{2D}; k, Re_{2D} \right) = \frac{1}{2} \int_{-1}^{1} \left| \sum^\infty_{j = 1}{ \left( A^{O}_{j} \hat{\hat{\eta}}^{O2D}_{j} \mbox{e}^{-i \omega^{O2D}_{j} t_{2D}} - \frac{i}{k}  B^{O}_{j} \hat{\eta}^{S2D}_{j} \mbox{e}^{-i \omega^{S2D}_{j} t_{2D}} \right) } \right|^{2} dy.
	\label{eq: integral_I2D}
\end{align}
The integral $\mathcal{I}_{2D} \left( t_{2D}; k, Re_{2D} \right)$ vanishes only at $t_{2D} = 0$ and is $\mathcal{O} \left(\mid \mbox{exp} \left( -2i \omega^{2D}_{max} t_{2D} \right) \mid\right)$, where $\omega^{2D}_{max}$ is the leading eigenvalue among $\omega^{O2D}_{j}$ and $\omega^{S2D}_{j}$ when $t_{2D}$ is large. Thus, the large time asymptotics using the extended Squire transformation imposes that
\begin{align}
G \left( t; \alpha, \beta, Re \right) \sim  \left( \frac{Re}{Re_{2D}} \right)^2 G_{2D} \left( t_{2D}; k, Re_{2D} \right),
\label{eq:G_asymp_largeRe_largeTime}
\end{align}
with
\begin{align}
G_{2D} \left( t_{2D}; k, Re_{2D} \right) = \operatorname*{sup}_{\forall \tilde{E_{k}(0)} \neq  0} \left[ \frac{\mathcal{I}_{2D} \left( t_{2D}; k, Re_{2D} \right)}{E_{k}(0)} \right].
\label{eq:G_asymp_largeRe_largeTime2}
\end{align}
Indeed, since $\mathcal{I}_{2D} \left( t_{2D}; k, Re_{2D} \right)$ is the function \eqref{eq: integral_I2D} independent of the coefficients $B^{S}_{j}$, it depends only on the initial wall-normal velocity and since, maximizing the gain imposes to minimize $E_{k}(0)$ at constant $\mathcal{I}_{2D}\left( t_{2D}; k, Re_{2D} \right)$, the initial wall-normal vorticity $\tilde{\eta}_{0}$ should be set to zero. The optimal for $G_{2D}$ should be searched only in the initial perturbations field $\tilde{v}_{0}(y)$ as in the previous case when $1 \ll t \ll {Re}/{Re_{2D}}$.

The extended Squire transformation, therefore, predicts according to the equations \eqref{eq:G_asymp_largeRe} and \eqref{eq:G_asymp_largeRe_largeTime} that, as soon as $t \gg 1$ (even if $t/Re \ll 1$), the entire optimal gain curve at large $Re$ ($\alpha \rightarrow 0$) is an unique curve dependent only on $Re_{2D}$ and $k$ given by $t_{2D}^{2} \mathcal{G}_{2D}$ at small $t_{2D}$ and $G_{2D} \left( t_{2D}; k, Re_{2D} \right)$ at $t_{2D}$ of order unity or large, once the gain is rescaled by $\left( Re_{2D}/Re \right)^{2}$ and the time by $\left( Re_{2D}/Re \right)$. It also implies that the optimal initial perturbations for optimization time $t$ large or $t_{2D}$ arbitrary (small or large) involve only $\tilde{v}_{0}(y)$ component i.e. $\tilde{\eta}_{0}(y) = 0$. As we will see in \S\ref{sec:Discussion}, this result may be seen as an extension and an alternative formal proof of the classical scaling argument put forward by \cite{Gustavsson_1991, Reddy_n_Henningson_1993}.
%from the extended Squire transformation is extremely powerful since transient growth of nearly longitudinal $3D$ perturbations at large Reynolds numbers are predicted by the rescaled wall-normal vorticity of the $2D$ $OS$-modes.

\section{The Squire transformation extended to the Adjoint problem}
\label{sec: consequences_AjointProb}
The optimal gain can be analysed in a different limit, i. e. for finite $Re$ but as time $t$ goes to infinity, by introducing the adjoint equations with respect to the scalar product associated with the energy norm $\langle \hat{\textbf{\textit{q}}}_1,\ \hat{\textbf{\textit{q}}}_2 \rangle =	\int_{-1}^{1}{\hat{\textbf{\textit{q}}}_2^H \textbf{M} \hat{\textbf{\textit{q}}}_1 dy}$, where $^H$ represents the conjugate-transpose of a matrix and $\textbf{M} = 
	k^{-2}\begin{bmatrix}
		(k^2 -D^2)	&0\\
		0 &1
	\end{bmatrix}$. The norm with respect to this scalar product is related to the kinetic energy defined in equation \eqref{eq: norm_KE} as  $E_{k}(t) = \frac{1}{2}\left\| \tilde{\textbf{\textit{q}}} \right\|^2$. Thus, the adjoint equations are
\begin{equation}
	-\frac{\partial}{\partial t}
		\begin{bmatrix}
			k^2-D^2	&0\\
			0	&1
		\end{bmatrix} 
	\tilde{\textbf{\textit{q}}}^{\dagger} = 
	\begin{bmatrix}
		L^{O\dagger}	&-i\beta \frac{dU_0}{dy}\\
		0 &L^{S\dagger}
	\end{bmatrix}
	\tilde{\textbf{\textit{q}}}^{\dagger},
\label{eq:LOSSQadjoint_matrix}
\end{equation}
where $L^{O\dagger}$ and $L^{S\dagger}$ represent the adjoint Orr-Sommerfeld and Squire operators, respectively,
\begin{align}
	L^{O\dagger} = -i\alpha U_0 \left( k^2-D^2 \right) + 2i\alpha \frac{dU_0}{dy}D + \frac{1}{Re}\left( k^2-D^2 \right)^2,
	\label{eq:L_adjoint_OS1}
\end{align}
\begin{align}
	L^{S\dagger} =  -i\alpha U_0 + \frac{1}{Re}\left( k^2-D^2 \right),
	\label{eq:L_adjoint_SQ1}
\end{align}
and the adjoint state vector is $\tilde{\textbf{\textit{q}}}^{\dagger}$ = $[\tilde{v}^{\dagger}\left(y,t; \alpha, \beta, Re\right), \tilde{\eta}^{\dagger}\left(y,t; \alpha, \beta, Re\right)]^T$. Here, $\tilde{v}^{\dagger}(y,t;$ $\alpha, \beta, Re)$ and $\tilde{\eta}^{\dagger}\left(y,t; \alpha, \beta, Re\right)$ denote the adjoint wall-normal velocity and vorticity components, respectively. The spectrum of the adjoint $OS$ operator $L^{O\dagger}$ is the complex conjugate of the spectrum of the direct $OS$-operator $L^{O}$ and similarly for the adjoint $SQ$-operator $L^{S\dagger}$. But in the adjoint linear operator \eqref{eq:LOSSQadjoint_matrix}, it is the wall-normal vorticity $\tilde{\eta}^{\dagger}$ that forces the adjoint wall-normal velocity equation whereas the adjoint Squire equation is independent of the adjoint wall-normal velocity $\tilde{v}^{\dagger}_{j}$. The adjoint $OS$-modes $\hat{\textbf{\textit{q}}}^{O\dagger}_{j} = \left[ \hat{v}^{O\dagger}_{j},  0 \right]^{T}$ correspond then to zero wall-normal vorticity and the adjoint $SQ$-modes $\hat{\textbf{\textit{q}}}^{S\dagger}_{j} = \left[ \hat{v}^{S\dagger}_{j},  \hat{\eta}^{S\dagger}_{j} \right]^{T}$ have a non-zero wall-normal velocity corresponding to the forcing of the adjoint $OS$ operator by the off-diagonal term $-i\beta \frac{dU_0}{dy} \hat{\eta}^{S\dagger}_{j}$ in the adjoint equation \eqref{eq:LOSSQadjoint_matrix}. 

The Squire transformation also applies to the homogeneous part of the adjoint Orr-Sommerfeld equation and to the adjoint Squire equation. Thus, a $3D$ adjoint $OS$-mode at any $\alpha$, $\beta$ and $Re$, is related to a $2D$ adjoint $OS$-mode at $\alpha_{2D} = k$, $\beta_{2D} = 0$ and $Re_{2D}$ via the transformation
\begin{eqnarray}
\omega_{j}^{O*}(\alpha, \beta, Re) = \frac{\alpha}{k} \mbox{ } \omega_{j}^{O2D*}\left( k, Re_{2D} \right),
\label{eq:transform_lambda_adj}
\end{eqnarray}
\begin{eqnarray}
\hat{v}^{\dagger}_{j}(y; \alpha, \beta, Re) = \hat{v}^{O2D\dagger}_{j}(y; k, Re_{2D}).
\label{eq:sq_transform_v_adj1}
\end{eqnarray}
Similarly, the adjoint $SQ$-mode at any $\alpha$, $\beta$ and $Re$ reads
\begin{eqnarray}
\omega^{S*}_{j}(\alpha, \beta, Re) = \frac{\alpha}{k} \mbox{ } \omega^{S2D*}_{j}\left( k, Re_{2D} \right),
\label{eq:transform_mu_adj}
\end{eqnarray}
\begin{eqnarray}
\hat{\eta}^{S^\dagger}_{j}(y; \alpha, \beta, Re) = \hat{\eta}^{S2D\dagger}_{j}(y; k, Re_{2D}),
\label{eq:sq_transform_eta_adj}
\end{eqnarray}
\begin{eqnarray}
\hat{v}^{S\dagger}_{j}(y; \alpha, \beta, Re) = \beta \frac{k}{\alpha}\hat{v}^{S2D\dagger}_{j}(y; k, Re_{2D}),
\label{eq:sq_transform_v_adj}
\end{eqnarray}
where $\hat{v}^{S2D\dagger}_{j}$ is the rescaled wall-normal velocity that satisfies the two-dimensional adjoint Orr-Sommerfeld equation forced at the complex frequency $\omega^{S2D*}_{j}$  by the adjoint $SQ$-modes such that
\begin{align}
\left[i (\omega^{S2D*}_{j} + k U_0) (k^2-D^2) - 2ik \frac{dU_0}{dy}D - \frac{1}{Re_{2D}}(k^2-D^2)^2\right] \hat{v}^{S2D \dagger}_{j}\left(y; k, Re_{2D}\right) \notag \\ = -i \frac{dU_0}{dy} \hat{\eta}^{S2D \dagger}_{j}\left(y; k, Re_{2D}\right).
\label{eq:2D_adj_OS_eqn}
\end{align}
Thus, the Squire transformation extended to the adjoint modes predicts that the adjoint Squire mode should have a $\hat{v}^{\dagger}$-component scaling like ${Re}/{Re_{2D}}$.
\section{Consequences on long-time optimal gains}
\label{sec: consequences_long_time}
Since the basis of direct modes is biorthogonal to the basis of adjoint modes \citep{Schmid_n_Henningson_2001}, the coefficients in the eigenfunction expansion \eqref{eq:eig_expn_full} of the initial--value problem \eqref{eq:LOSSQ_matrix} for the wave vector $\vec{k} = \left( \alpha, \beta \right)$ at $Re$, are given by:
\begin{eqnarray}
A^{O}_{j} = \frac{\langle 
\tilde{\textbf{\textit{q}}}_{0},
\hat{\textbf{\textit{q}}}^{O\dagger}_{j} \rangle}{\langle 
\hat{\textbf{\textit{q}}}^{O}_{j},
	\hat{\textbf{\textit{q}}}^{O\dagger}_{j} \rangle}
	& \mbox{ and }
&A^{S}_{j} = \frac{\langle 
\tilde{\textbf{\textit{q}}}_{0},
\hat{\textbf{\textit{q}}}^{S\dagger}_{j} \rangle}{\langle 
\hat{\textbf{\textit{q}}}^{S}_{j},
	\hat{\textbf{\textit{q}}}^{S\dagger}_{j} \rangle},
\label{eq:coeff_AjandBj}
\end{eqnarray}
where $A^{S}_{j} = \left(\frac{\beta Re}{Re_{2D}}B^{O}_{j} + B^{S}_{j}\right)$. For $t \gg \left(\Delta \omega_{max} \right)^{-1}$, where $\Delta \omega_{max}$ is the difference in the growth rate of the first and the second leading eigenmode,  the long-time response is dominated by the leading eigenmode with a non-zero co-efficient in the solution \eqref{eq:eig_expn_full}.

Consider the case where the leading mode is the $OS$-mode $\hat{\textbf{\textit{q}}}^{O}_{1} = [\hat{v}^{O}_{1}, \hat{\eta}^{O}_{1}]^T$, then at $t \gg \left(\Delta \omega_{max} \right)^{-1}$, $\tilde{\textbf{\textit{q}}} \left( t \right) \sim {A^{O}_{1} \hat{\textbf{\textit{q}}}^{O}_{1}	\exp{\left( -i\omega^{O}_{1}t\right)}}$ and the optimization problem for long-time gain reduces to maximizing the coefficient $A^{O}_{1}$. Expression \eqref{eq:coeff_AjandBj} shows classically that the large-time gain is achieved by taking the leading adjoint $OS$-mode $\hat{\textbf{\textit{q}}}^{O\dagger}_{1} = [\hat{v}^{O\dagger}_{1}(y), 0]^T$ as the initial condition. Hence, the gain reads
\begin{align}
G(\alpha, \beta, t; Re) \sim G^{O}_{\infty} \left|\mbox{e}^{-2i\omega^{O}_{1}t}\right| \mbox{ with } G^{O}_{\infty} = \frac{\left\| \hat{\textbf{\textit{q}}}^{O}_{1} \right\|^2 \left\| \hat{\textbf{\textit{q}}}^{O\dagger}_{1} \right\|^2}{\left|\langle \hat{\textbf{\textit{q}}}^{O}_{1},\ \hat{\textbf{\textit{q}}}^{O\dagger}_{1} \rangle\right|^2},
\label{eq:trans_gwth_long_time_OS}
\end{align}
where $G^{O}_{\infty}(\alpha, \beta, t; Re)$ is the extra gain compared to the exponential variation. Similarly, if the leading eigenmode is the $SQ$-mode $\hat{\textbf{\textit{q}}}^{S}_{1} = [0, \hat{\eta}_{1}(y)]^T$, $G^{S}_{\infty} = {\left\| \hat{\textbf{\textit{q}}}^{S}_{1} \right\|^2 \left\| \hat{\textbf{\textit{q}}}^{S\dagger}_{1} \right\|^2}/{\mid\langle \hat{\textbf{\textit{q}}}^{S}_{1},\ \hat{\textbf{\textit{q}}}^{S\dagger}_{1} \rangle \mid^2}$ denotes the extra gain compared to the exponential growth, or decay $\mid\mbox{e}^{-2i\omega^{S}_{1}t}\mid$.

When $\alpha \neq 0$, the extended Squire transformation states, as demonstrated in \S\ref{sec: consequences_eigen_fn_sqtrans} $\&$  \S\ref{sec: consequences_AjointProb} , that, for fixed $k = \sqrt{\alpha^2 + \beta^2}$ and $Re_{2D}$, the direct and the adjoint $OS$-modes transform as
\begin{align}
\hat{\textbf{\textit{q}}}^{O}_{j} = 
\begin{bmatrix}
			\hat{v}^{O2D}_{j}\\
%			\frac{\beta Re}{Re_{2D}} \hat{\eta}^{O2D}_{j}
			\left({i \beta Re}/{Re_{2D}}\right) \hat{\hat{\eta}}^{O2D}_{j}
\end{bmatrix}
\mbox{ and }
\hat{\textbf{\textit{q}}}^{O\dagger}_{j} = 
\begin{bmatrix}
			\hat{v}^{O2D\dagger}_{j}\\
			0
\end{bmatrix},
\label{eq:DirectAndAdjointOSmodeTransform}
\end{align}
with $\beta = k \sqrt{1 - {Re_{2D}^2}/{Re^2}}$. Therefore, according to the extended Squire transformation, the long-time extra gain $G^{O}_{\infty}\left(\alpha, \beta, Re\right)$ may be rewritten as a product of $2D$ and $3D$ long-time extra gains:
\begin{align}
G^{O}_{\infty}\left(\alpha, \beta, Re\right) = G^{O2D}_{\infty}\left( k, Re_{2D} \right) \left( 1 + \frac{\beta^2 Re^2}{Re_{2D}^2} G^{O3D}_{\infty}\left( k, Re_{2D} \right) \right),
\label{eq:trans_gwth_long_time_result_OS}
\end{align}
with $G^{O2D}_{\infty}\left( k, Re_{2D} \right)
 = {\left\| \hat{\textbf{\textit{q}}}^{O 2D}_{1} \right\|^2 \left\| \hat{\textbf{\textit{q}}}^{O 2D\dagger}_{1} \right\|^2}/{\mid\langle \hat{\textbf{\textit{q}}}^{O2D}_{1},\ \hat{\textbf{\textit{q}}}^{O2D\dagger}_{1} \rangle\mid^2}$ given by
\begin{align}
G^{O2D}_{\infty}\left( k, Re_{2D} \right) = \frac{\int^{1}_{-1}{ \left( \left|\hat{v}^{O2D}_{1}\right|^2 + k^{-2}\left|D\hat{v}^{O2D}_{1}\right|^2 \right)dy }  \int^{1}_{-1}{ \left( \left|\hat{v}^{O2D\dagger}_{1}\right|^2 + k^{-2}\left|D\hat{v}^{O2D\dagger}_{1}\right|^2 \right)dy }}{ \left| \int^{1}_{-1}{ \left( \hat{v}^{O2D\dagger*}_{1} \hat{v}^{O2D}_{1} + k^{-2} D\hat{v}^{O2D\dagger*}_{1} D\hat{v}^{O2D}_{1} \right)dy } \right|^2}
\label{eq:trans_gwth_long_time_result_OS2D}
\end{align}
and
\begin{align}
G^{O3D}_{\infty}\left( k, Re_{2D} \right) = \frac{ k^{-2} \int^{1}_{-1}{ \left|\hat{\hat{\eta}}^{O2D}_{1}\right|^2 dy }}{ \int^{1}_{-1}{ \left( \left|\hat{v}^{O2D}_{1}\right|^2 + k^{-2}\left|D\hat{v}^{O2D}_{1}\right|^2 \right)dy }},
\label{eq:trans_gwth_long_time_result_OS3D}
\end{align}
where all fields are evaluated for $k$ and $Re_{2D}$ and here (and also, hereafter), they were written without the explicit dependence for the sake of brevity. The $G^{O2D}_{\infty} \left(k, Re_{2D}\right)$ is the extra-gain that would be obtained in the $2D$ case and it is known to result from the classical Orr-mechanism. The term $\left( \beta^2 Re^2/Re_{2D}^2 \right) G^{O3D}_{\infty}\left(k, Re_{2D}\right)$ is the extra-gain from the $3D$-effect, the contribution to the optimal transient growth arising from the lift-up mechanism due to the forcing of the wall-normal vorticity by the wall-normal velocity. Furthermore, the extended Squire transformation explains the form of the $3D$ contribution $\left( 1 + {\beta^2 Re^2}/{Re_{2D}^2} G^{O3D}_{\infty}\left( k, Re_{2D} \right) \right)$ with $G^{O3D}_{\infty}\left(k, Re_{2D}\right)$ that depends only on $2D$ eigenfunctions $\hat{v}^{O2D}_{1}$ and $\hat{\hat{\eta}}^{O2D}_{1}$ introduced in \S\ref{sec: consequences_eigen_fn_sqtrans}. Contrary to the previous section where equations \eqref{eq:G_asymp_largeRe} and \eqref{eq:G_asymp_largeRe_largeTime} were the large Reynolds number asymptotic for the gain curve valid for all times via the extended Squire transform, the present prediction \eqref{eq:trans_gwth_long_time_result_OS} is valid for arbitrary Reynolds number but only for large time ${t Re_{2D}}/{Re} \gg 1$.

Similarly, for the direct and adjoint Squire modes the extended Squire transformation, as already demonstrated, gives
\begin{align}
\hat{\textbf{\textit{q}}}^{S}_{j} = 
\begin{bmatrix}
			0\\
			\hat{\eta}^{S2D}_{j}
\end{bmatrix}
\mbox{ and }
\hat{\textbf{\textit{q}}}^{S\dagger}_{j} = 
\begin{bmatrix}
			{\beta Re}/{Re_{2D}} \hat{v}^{S2D\dagger}_{j}\\
			\hat{\eta}^{S2D\dagger}_{j}
\end{bmatrix}.
\label{eq:DirectAndAdjointSQmodeTransform}
\end{align}
Using this, the long-time extra-gain can be rewritten as
\begin{align}
G^{S}_{\infty}(\alpha, \beta, Re) = G^{S2D}_{\infty}\left( k, Re_{2D} \right) \left( 1 + \left( \frac{\beta Re}{Re_{2D}} \right)^2 G^{S3D}_{\infty}\left( k, Re_{2D} \right) \right),
\label{eq:trans_gwth_long_time_result_SQ}
\end{align}
where
\begin{align}
G^{S2D}_{\infty} \left( k, Re_{2D} \right) = \frac{\int^{1}_{-1}{ \left|\hat{\eta}^{S2D}_{1}\right|^2dy }  \int^{1}_{-1}{ \left|\hat{\eta}^{S2D\dagger}_{1}\right|^2dy }}{ \left| \int^{1}_{-1}{ \hat{\eta}^{S2D\dagger*}_{1} \hat{\eta}^{S2D}_{1}dy}\right|^2}
\label{eq:trans_gwth_long_time_result_SQ2D}
\end{align}
is the $2D$ extra-gain for the $2D$ Squire mode which will be found numerically (see the results discussed in the next section) to be close to unity for all $k$ and $Re_{2D}$. The rescaled contribution $G^{S3D}_{\infty}$ corresponds to the lift-up phenomenon when seen as an initial--value given by the adjoint $SQ$-mode which has a $\beta Re/Re_{2D}$ larger $\hat{v}$-component than the $\hat{\eta}$-component:
\begin{align}
G^{S3D}_{\infty}\left( k, Re_{2D} \right) = \frac{ \int^{1}_{-1}{ \left( \left|\hat{v}^{S2D\dagger}_{1}\right|^2 + k^{-2}\left|D\hat{v}^{S2D\dagger}_{1}\right|^2 \right)dy } }{ k^{-2} \int^{1}_{-1}{ \left|\hat{\eta}^{S2D\dagger}_{1}\right|^2 dy }}.
\label{eq:trans_gwth_long_time_result_SQ3D}
\end{align}

Thus, the extra-gain for both $OS$ and $SQ$ modes exhibits a lift-up contribution scaling like $\beta^2 Re^2/Re_{2D}^2$ when $Re \gg 1$ (note that $\beta = k \sqrt{1 - \left( Re_{2D}/ Re \right)^2}$ and $\alpha = k Re_{2D}/ Re$ in the Squire transformation). No matter if the $OS$-mode or $SQ$-mode is the least stable eigenmode, only the $OS$-mode exhibits a $2D$ extra-gain due to the Orr-mechanism that, as we shall see, explains why this mode determines the maximum transient growth.

\section{Discussion}
\label{sec:Discussion}

\subsection{Direct computations of optimal growth in plane Poiseuille and Couette flows}
\label{subsec:DiscussionShearFlows}
\begin{figure}
\psfrag{x}{\large{$\frac{t}{Re}$}}
\psfrag{y}{\large{G}}
\psfrag{y1}{$G^{O}_{\infty}$}
\psfrag{y2}{$G^{S}_{\infty}$}
\psfrag{b1}{$\alpha = 1$}
\psfrag{b2}{$\alpha = 0.707$}
\psfrag{b3}{$\alpha = 0.1$}
\psfrag{b4}{$\alpha = 10^{-2}$}
\psfrag{b5}{$\alpha = 10^{-3}$}
\psfrag{b6}{$\alpha = 0$}
\begin{center}
\epsfig{file=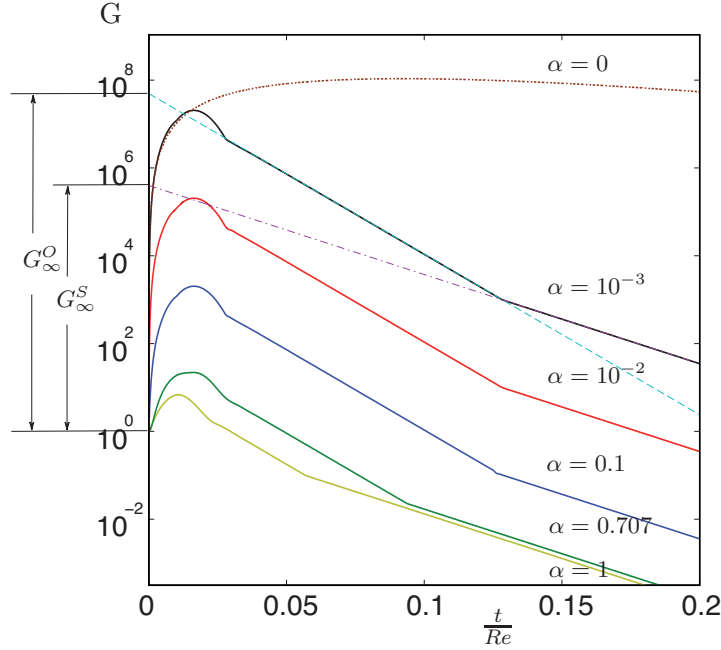,width=0.7\textwidth,keepaspectratio=true}
\end{center}
\caption{(colour online) Direct computations of optimal gain curves $G$ (solid lines) as a function of ${t}/{Re}$ in plane Poiseuille flow at $k = 1$ and $Re_{2D} = 1000$ for $\alpha = 1$, $0.707$, $0.1$, $10^{-2}$ and $10^{-3}$ corresponding respectively, via the Squire transform, to $Re = Re_{2D} = 1000$, $Re = 1414.2$, $Re = 10^{4}$, $Re = 10^{5}$ and $Re = 10^{6}$. The optimal gain curve for $\alpha = 0$ and $Re = 10^6$ is also presented ($\cdots$). For $\alpha = 10^{-3}$ (corresponding to $Re = 10^{6}$), the long-time exponential decay of the leading $OS$-mode $(---)$ and leading $SQ$-mode $(-\cdot-\cdot)$ are also displayed; they intersect the $y$-axis at $G^{O}_{\infty}$ and $G^{S}_{\infty}$, respectively, as given exactly by the equations \eqref{eq:trans_gwth_long_time_result_OS} and \eqref{eq:trans_gwth_long_time_result_SQ}.}
\label{fig:G_Vs_tbyRe}
\end{figure}

\begin{figure}
\psfrag{x}{\large{$\frac{t}{Re}$}}
\psfrag{y}{\large{$G\left( \frac{Re_{2D}}{\beta Re} \right)^2$}}
\psfrag{x1}{$\times 10^{-3}$}
\psfrag{b1}{$\alpha = 0.707$}
\psfrag{b2}{$\alpha = 0.1$}
\psfrag{b3}{$\alpha = 10^{-2}$}
\psfrag{b4}{$\alpha = 10^{-3}$}
%\psfrag{b5}{$\alpha = 0.9949$}
\begin{center}
\epsfig{file=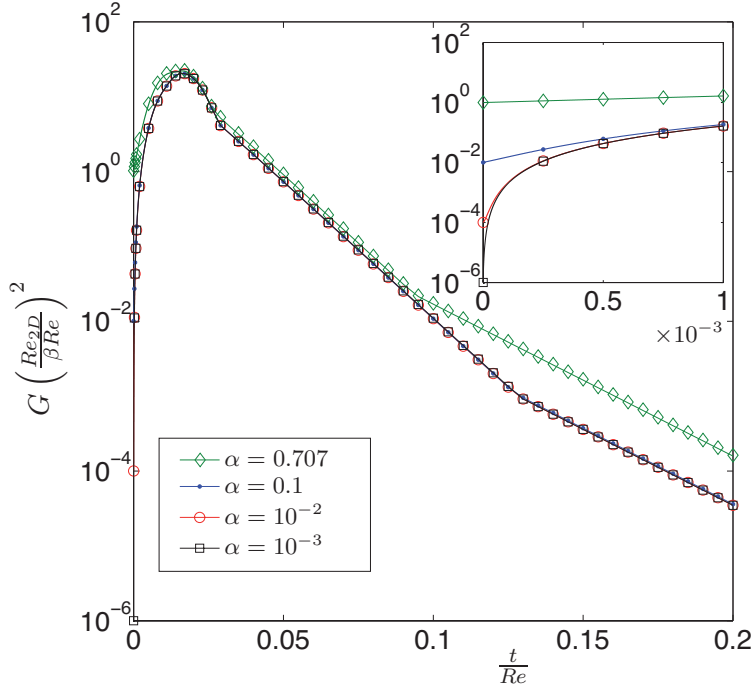,width=0.7\textwidth,keepaspectratio=true}
\end{center}
\caption{(colour online) Same data as in figure \ref{fig:G_Vs_tbyRe} but rescaled according to the large-$Re$ prediction via the extended Squire transformation for the optimal gain curve at ${t}/{Re}$ finite or large. Note that the curve $\alpha = 1$ (or $\beta = 0$) cannot be plotted in the present scaling. All the cases when $\alpha = 0.1$, $\alpha = 10^{-2}$ and $\alpha = 10^{-3}$ collapse so well for all times that they form a single curve; only in the inset, where the very early instants are shown, does a difference is visible since all curves should start at ${Re_{2D}^2}/{\beta^2 Re^2}$ for $t = 0$. But even in the close-up plot, the curves for $\alpha = 10^{-2}$ and $\alpha = 10^{-3}$ are indistinguishable except at the very first point at $t = 0$. The curve for $\alpha =  0.707$ is also very close to the large-$Re$ asymptotic curve and it only departs at large time.}
\label{fig:GtimesRe2D_by_betaRe_Vs_tbyRe}
\end{figure}

\begin{figure}
\psfrag{x}{\large{$\frac{t}{Re}$}}
\psfrag{y}{\large{G $\left( \frac{Re_{2D}}{\beta Re} \right)^2$}}
\psfrag{x1}{$\times 10^{-3}$}
\psfrag{x2}{$OS$--mode}
\psfrag{x3}{$SQ$--mode}
\psfrag{b1}{$\alpha = 0.707$}
\psfrag{b2}{$\alpha = 0.1$}
\psfrag{b3}{$\alpha = 10^{-2}$}
\psfrag{b4}{$\alpha = 10^{-3}$}
%\psfrag{b5}{$\alpha = 0.9949$}
\begin{center}
\epsfig{file=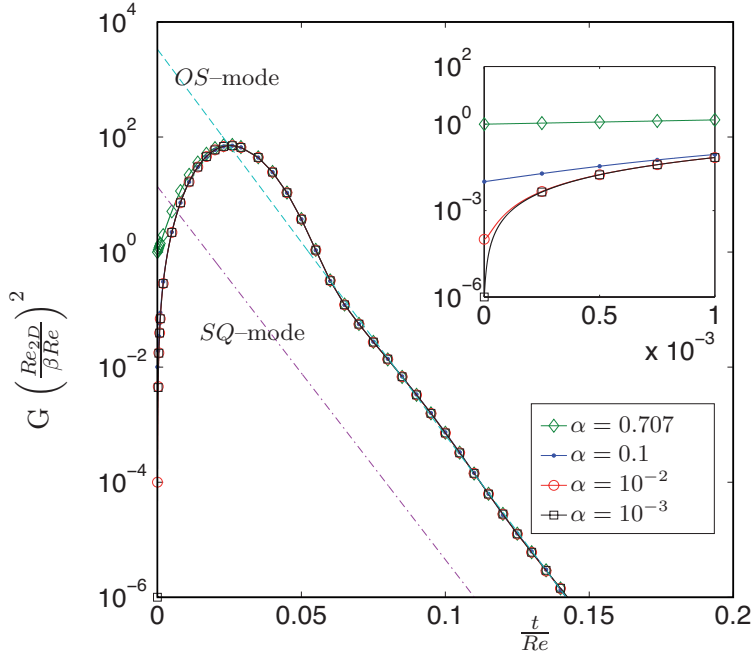,width=0.7\textwidth,keepaspectratio=true}
\end{center}
\caption{(colour online) Same as figure \ref{fig:GtimesRe2D_by_betaRe_Vs_tbyRe} but for the case of plane Couette flow ($k = 1$ and $Re_{2D} = 1000$). The long-time exponential decay of the leading $OS$-mode $(---)$ and leading $SQ$-mode $(-\cdot-\cdot)$ for the case of $\alpha = 10^{-3}$ (corresponding to $Re = 10^{6}$) are also displayed; they intersect the $y$-axis at $G^{O}_{\infty} \left( Re_{2D}^{2}/(\beta^2 Re^2) \right)$ and $G^{S}_{\infty} \left( Re_{2D}^{2}/(\beta^2 Re^2) \right)$, respectively.}
\label{fig:GtimesRe2D_by_betaRe_Vs_tbyReCouette}
\end{figure}

\begin{figure}
\psfrag{x}{\large{$\frac{\beta Re}{Re_{2D}}$}}
\psfrag{y}{\large{$G_{\infty}$}}
\psfrag{re}{$Re_{2D}$}
\psfrag{re1}{$10^{2}$}
\psfrag{re2}{$10^{3}$}
\psfrag{re3}{$5800$}
\psfrag{re4}{$5 \mbox{ } 10^{4}$}
\psfrag{y1}{$G^{O}_{\infty}$}
\psfrag{y2}{$G^{S}_{\infty}$}
\psfrag{y3}{\eqref{eq:trans_gwth_long_time_result_OS}}
\psfrag{y4}{$G^{S}_{\infty}$}
\psfrag{y5}{$G_{max}$}
\psfrag{y6}{\large{$G_{\infty}, G_{max}$}}
\psfrag{t1}{Lift-up}
\begin{center}
\epsfig{file=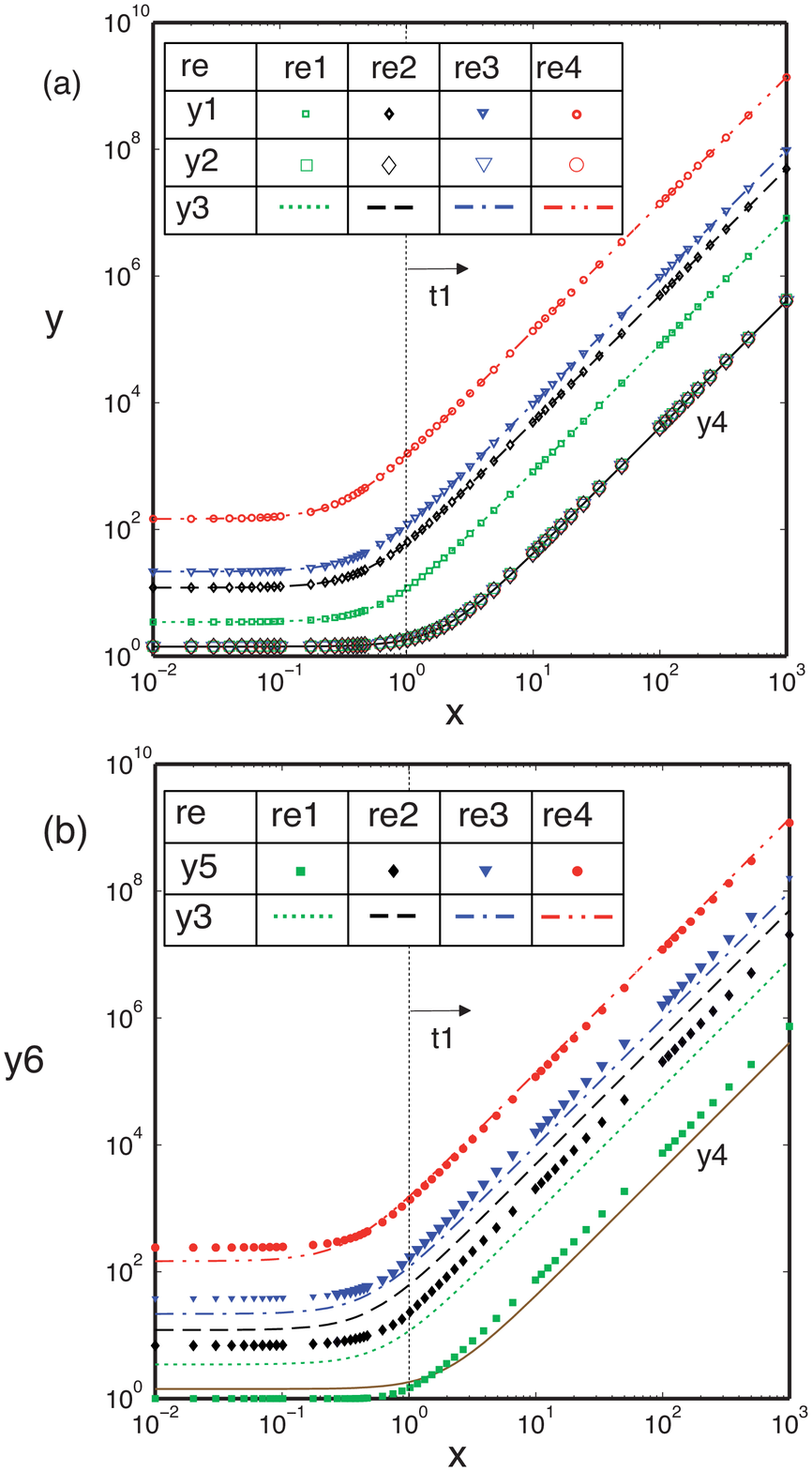,width=0.7\textwidth,keepaspectratio=true}
\end{center}
\caption{(a) (colour online) Long-time asymptotic predictions of the extended Squire transformation on the extra gain for all Reynolds numbers in plane Poiseuille flow is presented here by comparing results obtained via direct computations of the optimal long-time gains $G^{O}_{\infty}$ (small symbols) and $G^{S}_{\infty}$ (large symbols) as in figure \ref{fig:G_Vs_tbyRe} against the analytical formulae \eqref{eq:trans_gwth_long_time_result_OS} and \eqref{eq:trans_gwth_long_time_result_SQ} for $G^{O}_{\infty}$ (broken lines) and $G^{S}_{\infty}$ (solid line), respectively, when $k = 1$. The prediction of \eqref{eq:trans_gwth_long_time_result_SQ} is represented by the same solid line since $G^{S2D}_{\infty}$ and $G^{S3D}_{\infty}$ are identical at all $Re_{2D}$ considered (see table \ref{tab:table_Ginf}). (b) Comparison between the maximum optimal growth $G_{max}$ (closed symbols) and the optimal long-time gains $G_{\infty}$ over various $Re_{2D}$. Both $G_{\infty}$ and $G_{max}$ curves show the same trend but $G_{max}$ is approximately given by $G^{O}_{\infty}$ at large $Re_{2D}$.}
\label{fig:G_OS_n_SQ_Vs_betaRebyRe2D}
\end{figure}
Figure \ref{fig:G_Vs_tbyRe} displays optimal growth curves $G(t)$ (solid lines) directly computed using Singular Value Decomposition ($SVD$) as in \cite{jerome_chomaz_huerre}, for various Reynolds numbers $Re$ and wavenumbers $(\alpha, \beta)$ corresponding to the same $Re_{2D} = 1000$ and $k = 1$. The analytical predictions of the optimal long-time gains $G^{O}_{\infty} \mid \mbox{e}^{-2i\omega^{O}_{1}t}\mid$, $G^{S}_{\infty} \mid\mbox{e}^{-2i\omega^{S}_{1}t}\mid$ of the leading $OS$ and $SQ$ modes computed using the equations \eqref{eq:trans_gwth_long_time_result_OS} and \eqref{eq:trans_gwth_long_time_result_SQ} (dashed and dashed-dotted lines, respectively) at $Re = 10^6$ corresponding to $\alpha = 10^{-3}$ are also presented in the figure. The optimal growth at any time ${t}/{Re}$, increases with decreasing streamwise wavenumber $\alpha$ and after $t/Re \sim 0.03$, all optimal growth curves show two consecutive exponential decays (straight lines). In figure \ref{fig:G_Vs_tbyRe}, this two-step long-time dynamics can be identified with exponential decay of the leading $OS$-mode and $SQ$-mode. Their corresponding long-time optimal gains increase as $Re$ increases and $\alpha$ decreases as predicted by the scaling laws obtained using the extended-Squire transformation in  \S\ref{sec: consequences_InitValProb} and \S\ref{sec: consequences_long_time}. The two-step long-time behaviour occurs because $G^{O}_{\infty}$ is larger than $G^{S}_{\infty}$, a property retrieved for all the cases studied. When the leading eigenmode is an $OS$-mode, it dominates the optimal dynamics for all times large than $0.03Re$ and the piecewise exponential decay is not observed. Whereas, when the leading eigenmode is a $SQ$-mode, the $OS$-mode dominates after $t = 0.03Re$ but, since it decays faster than the $SQ$-mode, it is superseded after some time ${\left(\mbox{log} G_{\infty}^{O} - \mbox{log} G_{\infty}^{S}\right)}/{\Delta \omega_{max}}$ leading to the two-step optimal gain curve displayed in figure \ref{fig:G_Vs_tbyRe}. On figure \ref{fig:G_Vs_tbyRe}, it is also plotted in dotted line, the optimal gain for the longitudinal mode $\alpha = 0$ which is to be compared with the curve for $\alpha = 10^{-3}$ at the same $Re = 10^{6}$. The short-time behaviour is identical but after $t = 0.02Re$, the two-curves split apart as the gain for strictly longitudinal mode keeps increasing for a much longer time, thereby depicting the singularity of the longitudinal modes. It is also observed that $G^{O}_{\infty}$, given by the product of long-time optimal gain corresponding to $2D$-Orr mechanism $G^{O 2D}_{\infty}$ and $3D$ optimal gain from the lift-up mechanism $\left( {Re_{2D}^2}/{\beta^2 Re^2} \right) G^{O 3D}_{\infty}$, is approximately the maximum optimal growth for all Reynolds number and wavenumber shown here.

Figure \ref{fig:GtimesRe2D_by_betaRe_Vs_tbyRe} presents the optimal gain curves of figure \ref{fig:G_Vs_tbyRe} but rescaled as $G \left( {Re_{2D}^2}/{\beta^2 Re^2} \right)$ in order to verify the predictions of the extended Squire transformation on the large-$Re$ limit for the optimal gain curve at all time derived in \S\ref{sec: consequences_InitValProb}. Note that this rescaled gain diverges for the $2D$-case (when $\beta = 0$) and hence this case is not shown in figure \ref{fig:GtimesRe2D_by_betaRe_Vs_tbyRe}. As $Re$ increases, the rescaled optimal gain curves remarkably collapse into a single curve. The convergence is so strong that even at ${Re}/{Re_{2D}} = 10$ (i. e. $Re = 10^{4}$, $\alpha = 0.1$), the large-$Re$ asymptote is reached for all ${t}/{Re}$ and at ${Re}/{Re_{2D}} = \sqrt{2}$ (corresponding to $\alpha = 0.7071$) the asymptotic curve is nearly achieved. Only at very small ${t}/{Re}$ shown in the inset a departure of the curve may be observed since the new Squire transformed gain is not valid at the very initial instant where it should converge to unity. This confirms the large Reynolds number asymtotics predicted by the Squire transformation on the initial--value problem (eqns. \eqref{eq:G_asymp_largeRe} and \eqref{eq:G_asymp_largeRe_largeTime}) for all times larger than unity ($t \gg 1$ but ${t}/{Re}$ small, order unity or larger). 

	The rescaled optimal gain $G \left( {Re_{2D}^2}/{\beta^2 Re^2} \right)$  for the case of plane Couette flow at the same $Re_{2D} = 1000$ and $k = 1$ is shown in figure \ref{fig:GtimesRe2D_by_betaRe_Vs_tbyReCouette}. The symbols correspond to the same Reynolds numbers $Re$ and streamwise wavenumber $\alpha$ as in figure \ref{fig:GtimesRe2D_by_betaRe_Vs_tbyRe}. The curves are indistinguishable for all $Re$ and $\alpha$, including $\alpha = 0.707$ corresponding to $Re = 1414.2$. Thus, figures \ref{fig:GtimesRe2D_by_betaRe_Vs_tbyRe} \& \ref{fig:GtimesRe2D_by_betaRe_Vs_tbyReCouette} show that the large Reynolds number scaling of optimal growth curves obtained from the extended Squire transformation in \S\ref{sec: consequences_InitValProb} is extremely efficient in predicting the entire optimal gain curve. Also displayed in figure \ref{fig:GtimesRe2D_by_betaRe_Vs_tbyReCouette} are the long-time exponential decay of the leading $OS$ and $SQ$ modes (denoted, respectively, by dashed and dash-dotted lines) for $\alpha = 10^{-3}$ (corresponding to $Re = 10^{6}$). At $Re_{2D} = 1000$ and $k = 1$, similar to the case of plane Poiseuille flow, the leading eigenmode is a $SQ$-mode \citep{Schmid_n_Henningson_2001} and the tail of the optimal gain curve (corresponding to $t>>1$) could be expected to show two exponential decay rate. But, in this case, the exponential decay rates of the leading $OS$ and $SQ$ modes differ only in the third significant digit. Thus, for the optimization times shown in figure \ref{fig:GtimesRe2D_by_betaRe_Vs_tbyReCouette}, the optimal gain curve displays only one exponential decay corresponding to the leading $OS$-mode.
	
	Note that, for the optimal growth $G(t)$, the large-$Re$ rescaling obtained from the extended Squire transformation is similar to that proposed by \cite{Gustavsson_1991} who deduced large-$Re$ number scaling law for maximum optimal gain in plane Poiseuille flow but, here, wall-normal vorticity rescaling comes out naturally from the extended Squire transformation. Moreover, it is illustrated by comparing the results of large Reynolds number asymptotics and direct computations that the extended Squire transformation works for the entire optimal growth curve at all time ${t}/{Re}$ small, order unity or large.

\begin{figure}
\psfrag{x}{\large{$\frac{\beta Re}{Re_{2D}}$}}
\psfrag{y}{\large{$G_{\infty}$}}
\psfrag{re}{$Re_{2D}$}
\psfrag{re1}{$500$}
\psfrag{re2}{$10^{3}$}
\psfrag{re3}{$10^{4}$}
\psfrag{re4}{$10^{5}$}
\psfrag{y1}{$G^{O}_{\infty}$}
\psfrag{y2}{$G^{S}_{\infty}$}
\psfrag{y3}{\eqref{eq:trans_gwth_long_time_result_OS}}
\psfrag{y4}{$G^{S}_{\infty}$}
\psfrag{y5}{$G_{max}$}
\psfrag{y6}{\large{$G_{\infty}, G_{max}$}}
\psfrag{t1}{Lift-up}
\begin{center}
\epsfig{file=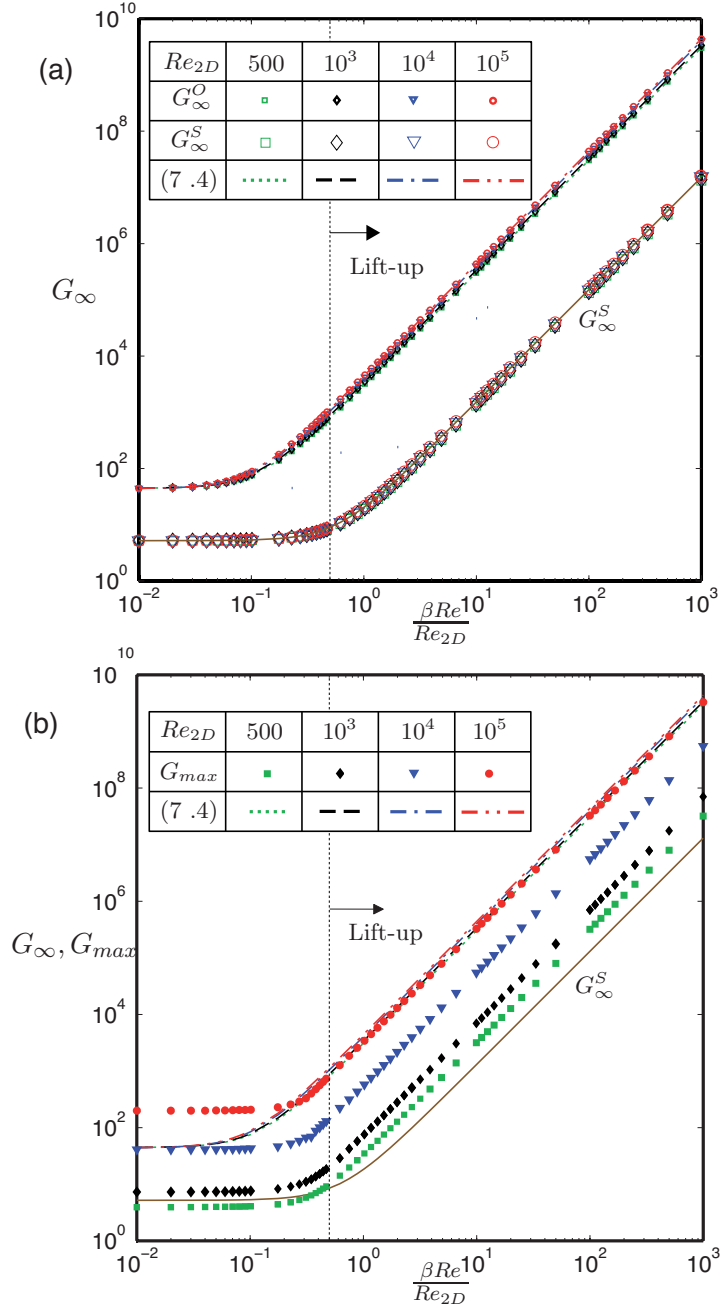,width=0.7\textwidth,keepaspectratio=true}
\end{center}
\caption{(colour online) Same as figure \ref{fig:G_OS_n_SQ_Vs_betaRebyRe2D} but for the case of plane Couette flow ($k = 1$).}
\label{fig:G_OS_n_SQ_Vs_betaRebyRe2DCouette}
\end{figure}

The variation of the long-time optimal gains, namely, $G_{\infty}^{O}$ and $G_{\infty}^{S}$, for arbitrary Reynolds numbers $Re$ are plotted in figure \ref{fig:G_OS_n_SQ_Vs_betaRebyRe2D}(a). The curves are obtained  via the equations \eqref{eq:trans_gwth_long_time_result_OS} and \eqref{eq:trans_gwth_long_time_result_SQ} for the various $2D$-Reynolds number $Re_{2D} = 10^2$, $Re_{2D} = 10^3$ and $Re_{2D} = 10^4$ at $k = 1$. The large and small symbols represent the quantities $G^{S}_{\infty}$ and $G^{O}_{\infty}$, respectively, directly computed using $SVD$ in plane Poiseuille flow as in figure \ref{fig:G_Vs_tbyRe}. The long-time gains at all Reynolds number are precisely predicted by the analytical formulae \eqref{eq:trans_gwth_long_time_result_OS} and \eqref{eq:trans_gwth_long_time_result_SQ} for all Reynolds numbers. As already observed in figure \ref{fig:G_Vs_tbyRe}, $G^{O}_{\infty}$ is always larger than $G^{S}_{\infty}$ in plane Poiseuille flow. Both gains, however, increase with Reynolds number $Re$ and vary as $Re^2$ at large Reynolds numbers. It is observed that $G^{S}_{\infty}$ does not change with respect to the $2D$-Reynolds number in the range considered: $Re_{2D} = 10^2, 10^3$ and $10^4$. Similarly, in the case of plane Couette flow, figure \ref{fig:G_OS_n_SQ_Vs_betaRebyRe2DCouette}(a) compares the long-time optimal gains $G^{O}_{\infty}$ and $G^{S}_{\infty}$ obtained via \eqref{eq:trans_gwth_long_time_result_OS} and \eqref{eq:trans_gwth_long_time_result_SQ} with that directly computed using $SVD$ as in figure \ref{fig:G_Vs_tbyRe} over various $Re_{2D}$. Here, again the analytical formulae  \eqref{eq:trans_gwth_long_time_result_OS} and \eqref{eq:trans_gwth_long_time_result_SQ} predict exactly the long-time gains. Also, $G^{O}_{\infty}$ is always larger than $G^{S}_{\infty}$. However, unlike the case for plane Poiseuille flow, not only $G^{S}_{\infty}$ but also $G^{O}_{\infty}$ does not vary much for a wide range of $2D$-Reynolds number.

In figures \ref{fig:G_OS_n_SQ_Vs_betaRebyRe2D}(b) \&  \ref{fig:G_OS_n_SQ_Vs_betaRebyRe2DCouette}(b), the maximum optimal gain $G_{max}$ (closed symbols) obtained via $SVD$ is compared with the long-time optimal gains $G^{O}_{\infty}$ and $G^{S}_{\infty}$ (using \eqref{eq:trans_gwth_long_time_result_OS} and \eqref{eq:trans_gwth_long_time_result_SQ})  for various Reynolds numbers $Re$ at fixed $2D$-Reynolds numbers $Re_{2D}$. All the data are computed for $k = 1$. When $\beta = 0$, $G_{max}$ is precisely the maximum transient growth corresponding to the $2D$ Orr-mechanism. For a given $Re_{2D}$ and $k$, both figures \ref{fig:G_OS_n_SQ_Vs_betaRebyRe2D}(b) \&  \ref{fig:G_OS_n_SQ_Vs_betaRebyRe2DCouette}(b) show that this value of $G_{max}$ is approximately constant as long as  $\beta < 1/\sqrt{2}$ (or $\beta Re/Re_{2D} < 1$). However, when $\beta \rightarrow k$ (or $Re/Re_{2D} \gg 1$), $G_{max}$ increases steeply as $\left(Re/Re_{2D}\right)^{2}$. Note that at this regime $G_{max}$ corresponds to the $3D$ lift-up mechanism. When $G_{max}$ is compared with the corresponding long-time extra-gains $G_{\infty}$, it is seen that they follow the same trend with respect to $\beta Re/Re_{2D}$ in both plane Poiseuille and plane Couette flows. When $Re_{2D}$ is small, $G_{max}$ corresponding to the lift-up mechanism shows large deviations from $G_{\infty}^{O}$ at all $\beta Re/Re_{2D}$. However, as $Re_{2D} \gg 1$, $G_{max}$ seems to remarkably converge toward $G_{\infty}^{O}$ at large $\beta Re/Re_{2D}$. This result is important as it shows that, at the large Reynolds number limit, the optimal gain is predicted by $G^{O}_{\infty}$ and is therefore, the product of the $2D$ Orr-mechanism and a lift-up contribution as given by \eqref{eq:trans_gwth_long_time_result_OS}. This result is in accordance with \cite{farrell_1993} who showed, in viscous constant shear flows, that arbitrary $3D$ perturbations grow with a combination of the lift--up mechanism and the Orr mechanism of the wall normal velocity. Our results for both plane Poiseuille and plane Couette flow indicate that this amplification process can be universal. And the interaction of the Orr mechanism with the lift--up mechanism determines the optimal growth.

\subsection{Gustavsson's large-Reynolds number scaling}
\label{subsec:DiscussionGustavssonReddyHenningson}
\cite{Gustavsson_1991} studied the effect of wall-normal velocity forcing on the equation governing the wall-normal vorticity $\tilde{\eta}(y, t)$: the inhomogeneous Squire equation. In particular, \cite{Gustavsson_1991} analysed the initial-value problem of $\tilde{\eta}(y, t)$ alone when the initial wall-normal vorticity is zero i. e., $\tilde{\eta}_{0} = 0$ and the initial wall-normal velocity is an eigenfunction of the Orr-Sommerfeld equation \eqref{eq:L_direct_OS1} i. e., $\tilde{v}_{0} = \hat{v}^{O}_{j}$. While doing so, \cite{Gustavsson_1991} and later, \cite{Reddy_n_Henningson_1993} who as opposed to vorticity growth, directly computed the optimal energy growth in plane Poiseuille and plane Couette flows, obtained large-Reynolds number scaling for $G_{max}$ by rescaling the wall-normal vorticity as
\begin{eqnarray}
\tilde{\eta}(y, t; \alpha, \beta, Re) = \beta Re \mbox{ } \bar{\eta}(y, t/Re; k, \alpha Re).
\label{eq:GustavScaling}
\end{eqnarray}
Note that this is equivalent to the extended Squire transformation \S\ref{sec: consequences_eigen_fn_sqtrans}, however, in the case of \cite{Gustavsson_1991} and \cite{Reddy_n_Henningson_1993} this rescaling, introduces $(\beta Re)^2$ in the energy norm: 
\begin{eqnarray}
\left\| \tilde{\textbf{\textit{q}}} \right\|^2 = \frac{1}{2} \int_{-1}^{1} \left[ \left(  \left| \tilde{v} \right|^2 + \frac{1}{k^2}\left| D\tilde{v} \right|^2 \right) + (\beta Re)^2 \frac{1}{k^2} \left| \bar{\eta} \right|^2 \right] dy,
	\label{eq:norm_KE_GustavScaling}
\end{eqnarray}
which, at $Re \gg 1$, implies that the optimal growth is dominated by the wall-normal vorticity growth 
\begin{align}
G\left(t; \alpha, \beta, Re\right) \sim (\beta Re)^2 \operatorname*{sup}_{\forall \tilde{v}_0 \neq  0, \tilde{\eta}_0 = 0 } \left[ \frac{E_{\bar{\eta}}\left( t/Re; k, \alpha Re \right)}{E_{\tilde{v}}\left( 0 \right)} \right],
\label{eq:trans_gwth_defn_GustavScaling}
\end{align}
where
\begin{align}
E_{\bar{\eta}}\left( t/Re; k, \alpha Re \right) &= \frac{1}{2} \int_{-1}^{1}  \frac{1}{k^{2}} \left| \bar{\eta} \right|^2 dy,
\label{eq:Evort_GustavScaling}\\
E_{\tilde{v}}\left( 0 \right) &= \frac{1}{2} \int_{-1}^{1} \left(  \left| \tilde{v}_{0} \right|^2 + \frac{1}{k^{2}} \left| D\tilde{v}_{0} \right|^2 \right) dy.
\label{eq:Evel_GustavScaling}
\end{align}

In the present analysis, however, we have applied the Squire transformation on both the Orr-Sommerfeld and Squire eigenfunctions. In addition, the extended Squire transformation is used on the initial--value problem \eqref{eq:LOSSQ_matrix} for arbitrary initial conditions, in order to derive asymptotic solutions at $Re \gg 1$ and exact optimal gains at large-time with reported effect on the $2D$ Orr-mechanism and the $3D$ lift-up mechanism. Thus, the extended Squire transformation gives an alternative proof of the Gustavsson's scaling for arbitrary $\alpha Re$ as $Re \rightarrow \infty$.

\subsection{Extension to confined shear flows with destabilizing temperature gradient}
\label{subsec:DiscussionWithHeat}
\begin{figure}
\psfrag{x}{\large{$\frac{t}{Re}$}}
\psfrag{y}{\large{$G^{RB}$ $\left( \frac{Re_{2D}}{\beta Re} \right)^2$}}
\psfrag{x1}{$\times 10^{-3}$}
\psfrag{b1}{$\alpha = 0.707$}
\psfrag{b2}{$\alpha = 0.1$}
\psfrag{b3}{$\alpha = 10^{-2}$}
\psfrag{b4}{$\alpha = 10^{-3}$}
%\psfrag{b5}{$\alpha = 0.9949$}
\begin{center}
\epsfig{file=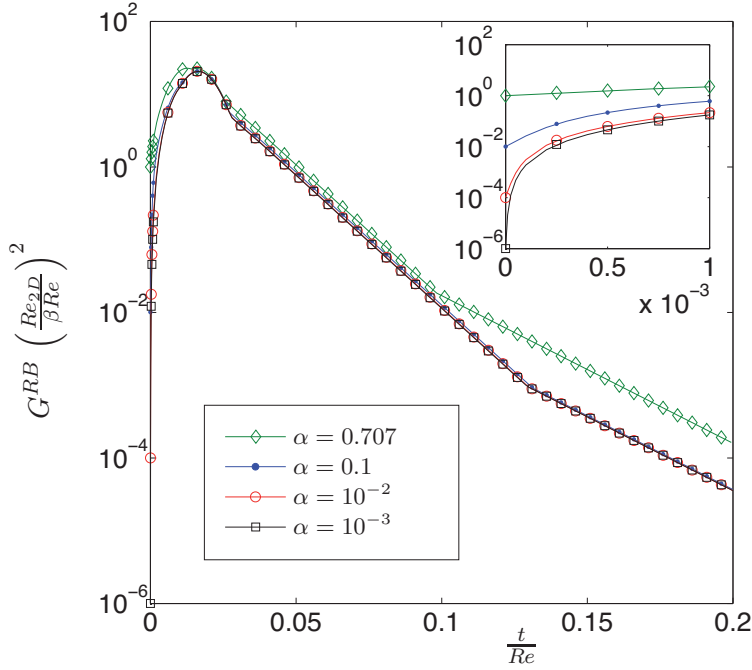,width=0.7\textwidth,keepaspectratio=true}
\end{center}
\caption{(colour online) Same as figure \ref{fig:GtimesRe2D_by_betaRe_Vs_tbyRe} but for the case of Rayleigh-B\'{e}nard-Poiseuille flow at Rayleigh number $Ra = 1000$ and Prandtl number $Pr = 1$. It shows that the large-Reynolds number asymptotic via the extended Squire transformation is also valid in shear flows with heat addition.}
\label{fig:THES_figureRBP}
\end{figure}

For the sake of simplicity, let us consider the so-called Rayleigh-B\'{e}nard-Poiseuille flow which is simply a channel flow with a constant temperature gradient (see for instance, \cite{Nicolas_2002, jerome_chomaz_huerre}). Nonetheless, the following analysis is true for arbitrary base flow temperature distributions. In general, the governing equations of the perturbation field \eqref{eq:LOSSQ_matrix} can be re-written in terms of the wall-normal velocity $\tilde{v}(y, t; \alpha, \beta, Re, Ra, Pr)$, temperature $\tilde{\theta}(y, t; \alpha, \beta, Re, Ra, Pr)$  and the wall-normal vorticity $\tilde{\eta}(y, t; \alpha, \beta, Re, Ra, Pr)$ at each wave vector $\vec{k} = \left( \alpha, \beta \right)^{T}$ \citep{Chandra_1961, Joseph_1976, Drazin_n_Reid_1981, jerome_chomaz_huerre}:
\begin{equation}
	-\frac{\partial}{\partial t} 
		\begin{bmatrix}
			k^2-D^2	&0	&0\\
			0	&1	&0\\
			0	&0	&1
		\end{bmatrix}
	 	\begin{bmatrix}
			\tilde{v}\\
			\tilde{\theta}\\
			\tilde{\eta}
		\end{bmatrix} 
	 = 
		\begin{bmatrix}
			L_{OS}					&-k^2 Ra/\left(Re^2 Pr\right) 	&0\\
			\frac{d\Theta_0}{dy} 	&L_{LHE}							&0\\
			i\beta \frac{dU_0}{dy}	&0 								&L_{SQ}\\
		\end{bmatrix}
		\begin{bmatrix}
			\tilde{v}\\
			\tilde{\theta}\\
			\tilde{\eta}
		\end{bmatrix},
	\label{eq:LOBE_matrix}
\end{equation}
with $D = \frac{\partial}{\partial y}$ and $k^2 = \alpha^2 + \beta^2$ as in the previous case. Here, $Ra =  {\alpha^* g  l^3 \Delta T}/{\nu^* \kappa^*}$ is the Rayleigh number and  $Pr = {\nu^*}/{\kappa^*}$ is the Prandtl numbers with $g$ the acceleration due to gravity, $\nu^*$ the kinematic viscosity, $\kappa^*$ the thermal diffusion coefficient and $\alpha^*$ the thermal expansion coefficient. Under the Boussinesq approximation, these parameters are functions of only $\Theta^*$, the average non-dimensional temperature of the channel (and hence, they do not depend on the temperature of the flow field). The space, time, velocity and temperature variables have been non-dimensionalized with respect to the characteristic length scale $l$, time scale $l/U$, velocity scale $U$ and temperature scale ${\Delta T}/{2}$, respectively. In the case of plane Poiseuille flow with constant cross-stream temperature gradient, $l$ is the half-channel width ${h}/{2}$, $U$ is the velocity at the centre of the channel and ${\Delta T}$ is the difference in temperature between the lower and upper wall. Equations \eqref{eq:LOBE_matrix} form the linearised Oberbeck-Boussinesq system of equations ($LOB$) wherein the operators $L_{OS}$ and $L_{SQ}$ are the usual Orr-Sommerfeld and Squire operators, given by \eqref{eq:L_direct_OS1} and \eqref{eq:L_direct_SQ1}. Whereas, the operator $L_{LHE}$ given by
\begin{align}
		L_{LHE} =  i\alpha U_0 + \frac{1}{Re Pr}\left( k^2-D^2 \right).
		\label{eq:L_direct_LHE}
\end{align}
comes from the linearized heat equation and it is the advection-diffusion operator governing the evolution of the temperature perturbation. These equations are to be solved for the boundary conditions: $\tilde{v}( \pm 1, t) = 0$, $D\tilde{v}(\pm 1, t) = 0$, $\tilde{\eta} (\pm 1, t) = 0$ and $\tilde{\theta}(\pm 1, t) = 0$. Here, the wall-normal velocity and temperature perturbations are coupled via the buoyancy terms whereas the wall-normal vorticity equation is decoupled from the temperature perturbations. The Squire equation is, however, forced by the solution of the coupled operator governing the wall-normal velocity and temperature perturbations.

For confined shear flows, the spectrum of \eqref{eq:LOBE_matrix} is discrete and complete \citep{Herron_1980} and it consists of two family of modes, namely, the Orr-Sommerfeld-Oberbeck-Boussinesq ($OSOB$) eigenfunctions $[ \hat{v}_j^{O},\hat{\theta}_j^{O}, \hat{\eta}_j^{O}]^{T}$ and the Squire ($SQ$) eigenfunctions $[ 0, 0, \hat{\eta}_j^{S}]^{T}$ with corresponding eigenvalues $\{\lambda_j^{O}\}$ and $\{\lambda_j^{S}\}$, respectively. They depend on $\alpha$, $\beta$, $Re$, $Ra$ and $Pr$. When $Ra > 1707.78$, the longitudinal $OSOB$-modes are destabilized as in the classical Rayleigh-B\'{e}nard convection.

For every given $Ra$ and $Pr$, the extended Squire transformation then relates oblique modes with $\alpha \neq 0$, $\beta \neq 0$ at Reynolds number $Re$ to a $2D$ spanwise-uniform mode with $\alpha_{2D} = k$, $\beta_{2D} = 0$ at a smaller Reynolds number $Re_{2D} =\left( \alpha/k \right) Re$:
\begin{align}
	\lambda_j^{O}(\alpha, \beta, Re, Ra, Pr) &= \frac{Re_{2D}}{Re}\lambda_j^{O2D}(k, Re_{2D}, Ra, Pr),
	\label{eq:SQ_transform1}\\
	\hat{v}_j^{O}(y; \alpha, \beta, Re, Ra, Pr) &= \hat{v}_j^{O2D}(y; k, Re_{2D}, Ra, Pr),
	\label{eq:SQ_transform2}\\
	\hat{\theta}_j^{O}(y; \alpha, \beta, Re, Ra, Pr) &= \frac{Re}{Re_{2D}}\hat{\theta}_j^{O2D}(y; k, Re_{2D}, Ra, Pr),
	\label{eq:SQ_transform3}\\
	\hat{\eta}_j^{O}(y; \alpha, \beta, Re, Ra, Pr)  &= \frac{i \beta Re}{Re_{2D}}\hat{\eta}_j^{O2D}(y; k, Re_{2D}, Ra, Pr),
	\label{eq:SQ_transform_OSOB}
\end{align}
in the case of the $OSOB$-modes and
\begin{align}
	\lambda_j^{S}(\alpha, \beta, Re, Ra, Pr) &= \frac{Re_{2D}}{Re}\lambda_j^{S2D}(k, Re_{2D}, Ra, Pr),
	\label{eq:SQ_transform4}\\
	\hat{\eta}_j^{S}(y; \alpha, \beta, Re, Ra, Pr)  &= \hat{\eta}_j^{S2D}(y; k, Re_{2D}, Ra, Pr),
	\label{eq:SQ_transform_SQRBP}
\end{align}
in the case of the $SQ$ modes. The superscripts $2D$ refer to variables of the $2D$ spanwise-uniform modes.

Using this transformation, the evolution of the perturbations in such flows can be written as
\begin{align}
\tilde{\textbf{\textit{q}}}\left(y, t; \alpha, \beta, Re, Ra, Pr\right)
=
\sum_j{
A^{O}_{j} \exp \left( -i\lambda_{j}^{O2D} Re_{2D}  \frac{t}{Re} \right)
\begin{bmatrix}
	\hat{v}_j^{O2D}\\
	\left({Re}/{Re_{2D}}\right) \hat{\theta}_j^{O2D}\\
	\left({i \beta Re}/{Re_{2D}}\right) \hat{\eta}_j^{O2D}\\
\end{bmatrix}
}
\notag \\+
\sum_j{ \left(\frac{i \beta Re}{Re_{2D}} B^{O}_{j} + B^{S}_{j} \right) \exp \left( -i\lambda_{j}^{S2D} Re_{2D}  \frac{t}{Re} \right)
\begin{bmatrix}
	0\\
	0\\
	\hat{\eta}_j^{S2D}\\
\end{bmatrix}}.
\label{eq:eig_expn_full_RBP}
\end{align}
%\end{widetext}
The Fourier amplitudes $\tilde{\textbf{\textit{q}}}\left(y, t; \alpha, \beta, Re, Ra, Pr\right) = [\tilde{v}, \tilde{\theta}, \tilde{\eta} ]^{T}$ are functions of $y$, $t$ and the control parameters, namely, $\alpha$, $\beta$, $Re$, $Ra$ and $Pr$. The coefficients $\{A^{O}_j\}$, $\{B^{O}_j\}$ and $\{B^{S}_j\}$ are complex constants that can be determined from the initial conditions on the state variables in the same manner as in the uniform temperature case, treated \eqref{eq:coeff_vel}, \eqref{eq:blaahhha} and \eqref{eq:coeff_eta}. 

Note that the wall-normal vorticity $\tilde{\eta}$ in the general solution  \eqref{eq:eig_expn_full_RBP} is given by exactly the same equation \eqref{eq:eig_expn_full_with_SQ_transform} and hence, it should obey the same scaling laws as in the uniform temperature case. Thus, all the results obtained in section \S\ref{sec: consequences_eigen_fn_sqtrans} apply equally well in such systems. 

The rescaled optimal gain $G_{RB} \left( Re_{2D}/{\beta Re} \right)^2$ curves at various Reynolds numbers for the case of Rayleigh-B\'{e}nard-Poiseuille flow is given in figure \ref{fig:THES_figureRBP}. The norm used to define the optimal gain $G_{RB}$ is taken as

\begin{align}
	\left\| \tilde{\textbf{\textit{q}}} \right\|^2 = \frac{1}{2} \int_{-1}^{1} \left[ \left| \tilde{v} \right|^2 + k^{-2} \left( \left| D\tilde{v} \right|^2 + \left| \tilde{\eta} \right|^2\right) \right] dy + \frac{1}{2} Ra \mbox{ } Pr \int_{-1}^{1} |\hat{\theta}|^2 dy,
	\label{eq: norm_RB}
\end{align}
since this choice for the relative weights of the thermal contribution to the energy is both coherent with the classical choice for the Rayleigh-B\'{e}nard problem in the absence of through flow and the classical potential energy for stably stratified flows (see \cite{jerome_chomaz_huerre} for details).
For the results displayed in figure \ref{fig:THES_figureRBP}, $Ra = 1000$ and $Pr = 1$ are taken along with $Re_{2D} = 1000$ and $k = 1$. The rescaled optimal gain curves are very similar to those in figure \ref{fig:GtimesRe2D_by_betaRe_Vs_tbyRe} corresponding to the uniform termperature case of plane Poiseuille flow. A perfect collapse is observed at all times $t/Re$ small, order unity or larger for $\alpha \leq 0.1$ or $Re \geq 10 Re_{2D}$. The mismatch occurs only for times $t/Re$ very small ($< 10^{-3}$) as shown in the inset of figure \ref{fig:THES_figureRBP}. This proves that the large-$Re$ number scaling \eqref{eq:G_asymp_largeRe} and \eqref{eq:G_asymp_largeRe_largeTime} derived via the extended Squire transformation are also applicable for confined shear flows with heat addition.

\section{Conclusion}
\label{sec:SQ_part_conclu}
The Squire transformation is extended to the wall-normal vorticity component of the Orr-Sommerfeld mode and the Squire mode. By introducing, two new fields for the wall-normal vorticity in the $2D$-case, any $3D$ eigenmode of the linearised Navier-Stokes equation is thus transformed into a three-component $2D$ eigenmode with $Re_{2D} = {\alpha Re}/{k}$ and $\alpha_{2D} = k$ in wall-bounded parallel flows. Consequently, as a manifestation of the lift-up mechanism, the wall-normal vorticity component in the $OS$-mode is transformed proportionally to the Reynolds number $Re$. In wall-bounded parallel flows, this extended Squire transformation allows us to solve the optimal gain at $t$ large but ${t}/{Re}$ arbitrary, for any large value of $Re$ with an exact renormalization of the entire gain curve depending only on $2D$ optimization.

	The Squire transformation is extended also to the adjoint eigenmodes. As a consequence, the optimal gain at large time $t \gg \left( \Delta \omega_{max} \right)^{-1}$, where $\Delta \omega_{max}$ is the difference between the first and second leading eigenmode growth rate, is expressed as an analytical function of $\beta^2 Re^2/Re_{2D}^{2}$ at a given $Re_{2D}$ and $k$ but arbitrary $Re$. If the leading eigenmode is an Orr-Sommerfeld mode, the large-time optimal gain at $t \gg \left( \Delta \omega_{max} \right)^{-1}$ is shown to be a product of respective gains from the $2D$ Orr-mechanism corresponding to $\hat{v}$-component of the $2D$ three-component $OS$-mode and the contribution of the $3D$ lift-up mechanism associated with the $\hat{\eta}$-component of the same mode.
	
	The results of these two asymptotic predictions (large $Re$ at arbitrary ${t}/{Re}$ and large $t$ but arbitrary $Re$, respectively) of the extended Squire transformation are verified for the case of plane Poiseuille flow, plane Couette flow and Rayleigh-B\'{e}nard-Poiseuille flow by direct numerical computations of optimal gain curves over a wide range of optimization time $t$. It is observed that, at large Reynolds numbers, the product of the gains from the $2D$ Orr mechanism and the lift-up mechanism is a good approximation to the maximum optimal transient growth.

J J S J thanks the financial support from the ``Direction des Relations Ext\'{e}rieures'' of \'{E}cole Polytechnique. The authors gratefully acknowledge Patrick Huerre, Cristobal Arratia and Yongyun Hwang for many fruitful discussion.

\begin{alphasection}

\iffalse
\section{Appendix}
\label{sec:Appendix}
Equations \eqref{eq:continuity_FourTransf} can lead to equations \eqref{eq:2Dcontinuity_FourTransf}, if 

\begin{align}
i k \hat{u}^{2D} = i \alpha \hat{u} + i \beta \hat{\mbox{w}},
\label{eq:u2d_sqtrans}
\end{align}
and
\begin{align}
\hat{v}^{2D} = \hat{v}.
\label{eq:v2d_sqtrans}
\end{align}
Under the classical Squire transformation, we have $Re_{2D} = \left( \alpha/k \right) Re$, $\omega_{2D} = \left( k/\alpha \right) \omega$ and the total wavelength $k$ is invariant. Thus, the evolution equation \eqref{eq:ymomentum_FourTransf} becomes independent of $\alpha, \beta$ and $Re$, if the pressure is transformed as
\begin{align}
\hat{p}^{2D} = \frac{k}{\alpha}\hat{p},
\label{eq:p2d_sqtrans}
\end{align}
and the evolution equation \eqref{eq:x2Dmomentum_FourTransf} for $\hat{u}^{2D}$ can be derived by manipulating \eqref{eq:xmomentum_FourTransf} and \eqref{eq:zmomentum_FourTransf} along with $\alpha, \beta, k$ and using \eqref{eq:u2d_sqtrans} \& \eqref{eq:p2d_sqtrans}. Finally, using \eqref{eq:u2d_sqtrans} \& \eqref{eq:p2d_sqtrans} in \eqref{eq:zmomentum_FourTransf} suggests that it becomes independent of $\alpha, \beta$ and $Re$ if
\begin{align}
\hat{\mbox{w}}^{2D} = \frac{k}{\beta}\hat{\mbox{w}}.
\label{eq:w2d_sqtrans}
\end{align}
It is then straight-forward to obtain \eqref{eq:u_sqtrans} from \eqref{eq:u2d_sqtrans} and \eqref{eq:w2d_sqtrans}.
\fi

\section{Annexe}
\label{sec:Annexe}
For the case of plane Poiseuille and plane Couette flows, table \ref{tab:table_Ginf} provides typical values of long-time optimal gains as obtained from  \eqref{eq:trans_gwth_long_time_result_OS} \& \eqref{eq:trans_gwth_long_time_result_SQ}. Here, $G^{S2D}_{\infty}$ and $G^{O2D}_{\infty}$ refer to the long-time optimal gains via $2D$ mechanisms corresponding to the leading Squire and Orr-Sommerfeld modes, respectively. Similarly, $G^{S3D}_{\infty}$ and $G^{O3D}_{\infty}$ refer to the long-time optimal gains via $3D$ lift-up mechanisms corresponding respectively to the leading Squire and Orr-Sommerfeld modes. As already seen in figures \ref{fig:G_OS_n_SQ_Vs_betaRebyRe2D} \& \ref{fig:G_OS_n_SQ_Vs_betaRebyRe2DCouette}, the long-time optimal gain $G^{S}_{\infty}$ is always less than $G^{O}_{\infty}$.

%%\begin{table}
%%  \begin{center}
%%   %\begin{minipage}{6cm}
%%    \begin{tabular}{c|cccc}
%%		%\cline{1-5} & & & &\\
%%		$Re_{2D}$& $10^{2}$& $10^{3}$& $5800$& $5 \cdot 10^{4}$ \\
%%		\cline{1-5} & & & &\\
%%		$G^{O2D}_{\infty}$& $3.4$& $12$& $21.6$& $146.7$ \\
%%		%\cline{1-5} & & & &\\
%%		$G^{S2D}_{\infty}$& $1.4$& $1.4$& $1.4$& $1.4$ \\
%%		%\cline{1-5} & & & &\\
%%		$G^{O3D}_{\infty}$& $2.4$& $4.1$& $4.5$& $9.4$ \\
%%		%\cline{1-5} & & & &\\
%%		$G^{S3D}_{\infty}$& $0.3$& $0.28$& $0.29$& $0.29$ \\
%%		%\cline{1-5}
%%	\end{tabular}
%%   %\end{minipage}
%%  \end{center}
%%\caption{Plane Poiseuille flow.} \label{tab:table_plane_Poiseuille_flow}
%%\end{table}
%%
%%\begin{table}
%%  \begin{center}
%%   %\begin{minipage}{6cm}
%%    \begin{tabular}{c|cccc}
%%		%\cline{1-5} & & & &\\
%%		$Re_{2D}$& $500$& $10^{3}$& $10^{4}$& $10^{5}$ \\
%%		\cline{1-5} & & & &\\
%%		$G^{O2D}_{\infty}$& $44.2$& $44.2$& $43.7$& $43.8$ \\
%%		%\cline{1-5} & & & &\\
%%		$G^{S2D}_{\infty}$& $5.2$& $5.2$& $5.2$& $5.2$ \\
%%		%\cline{1-5} & & & &\\
%%		$G^{O3D}_{\infty}$& $68.9$& $75.8$& $91.6$& $100$ \\
%%		%\cline{1-5} & & & &\\
%%		$G^{S3D}_{\infty}$& $2.5$& $2.6$& $2.8$& $2.9$ \\
%%		%\cline{1-5}
%%	\end{tabular}
%%   %\end{minipage}
%%  \end{center}
%%\caption{Plane Couette flow.} \label{tab:table_plane_Couette_flow}
%%\end{table}

\begin{table}
	\begin{center}
	\begin{tabular}{c| c c c c | c c c c }
			{} 		&\multicolumn{4}{c|}{{\textbf{plane Poiseuille flow}}}			&\multicolumn{4}{c}{{\textbf{plane Couette flow}}}\\
			\cline{1-9}
			\cline{1-9} & & & & & & &\\
		$Re_{2D}$& $10^{2}$& $10^{3}$& $5800$& $5 \cdot 10^{4}$& $500$& $10^{3}$& $10^{4}$& $10^{5}$ \\
			\cline{1-9} & & & & & & &\\
		$G^{O2D}_{\infty}$& $3.4$& $12$& $21.6$& $146.7$& $44.2$& $44.2$& $43.7$& $43.8$ \\
			\cline{1-9} & & & & & & &\\
		$G^{S2D}_{\infty}$& $1.4$& $1.4$& $1.4$& $1.4$& $5.2$& $5.2$& $5.2$& $5.2$ \\
			\cline{1-9} & & & & & & &\\
		$G^{O3D}_{\infty}$& $2.4$& $4.1$& $4.5$& $9.4$& $68.9$& $75.8$& $91.6$& $100$ \\
			\cline{1-9} & & & & & & &\\
		$G^{S3D}_{\infty}$& $0.3$& $0.28$& $0.29$& $0.29$& $2.5$& $2.6$& $2.8$& $2.9$ \\
		\end{tabular}
	\end{center}
\caption{Long-time optimal gains for plane Poiseuille and plane Couette flows at various Reynolds numbers $Re_{2D}$.} \label{tab:table_Ginf}
\end{table}

\end{alphasection}

\bibliographystyle{jfm}
\bibliography{the_Squire_transformation_and_3D_disturbances}
\end{document}